\documentclass[aps,onecolumn,groupedaddress,floatfix,longbibliography,superscriptaddress,notitlepage]{revtex4}

\usepackage{amsmath}
\usepackage{amssymb}
\usepackage{amsfonts}
\usepackage{graphicx}
\usepackage{color}
\usepackage{bm}
\usepackage{times,float}
\usepackage{graphicx}
\usepackage[usenames,dvipsnames,svgnames]{xcolor}
\usepackage{hyperref}
\hypersetup{colorlinks=true, linkcolor=magenta, citecolor=PineGreen,urlcolor=magenta}
\usepackage{fancyhdr}
\usepackage{soul}

\usepackage[caption=false]{subfig}
\usepackage{mathtools,leftidx}

\DeclareMathAlphabet\mathbfcal{OMS}{cmsy}{b}{n}

\begin{document}

\title{ Optical scalar beam propagation in nontrivial spacetime backgrounds }

\author{C. A. Escobar}
\email{carlos.escobar@xanum.uam.mx}
\affiliation{Av. Ferrocarril San Rafael Atlixco, Núm. 186, Col. Leyes de Reforma 1 A Sección, Alcaldía Iztapalapa, C.P. 09310, Ciudad de México}

\author{A. Mart\'{i}n-Ruiz}
\email{alberto.martin@nucleares.unam.mx}
\affiliation{Instituto de Ciencias Nucleares, Universidad Nacional Aut\'{o}noma de M\'{e}xico, 04510 Ciudad de M\'{e}xico, M\'{e}xico}

\author{M. Mondrag\'on}
\email{myriam@fisica.unam.mx}
\affiliation{Instituto de F\'{i}sica, Universidad Nacional Aut\'{o}noma de M\'{e}xico, Apartado Postal 20-364, Ciudad de M\'{e}xico 01000, M\'{e}xico}

\author{R. J\'auregui}
\email{rocio@fisica.unam.mx}
\affiliation{Instituto de F\'{i}sica, Universidad Nacional Aut\'{o}noma de M\'{e}xico, Apartado Postal 20-364, Ciudad de M\'{e}xico 01000, M\'{e}xico}

\begin{abstract}
We study the propagation of structured optical scalar beams in a spacetime background parameterized by a second-rank symmetric tensor. An analytic expression for the Green's function in a cylindrical coordinate system is obtained for particular choices of such a tensor. This facilitates the numerical exploration of the propagation of apertured Gaussian beams in this nontrivial background. Unusual focusing properties are found along with a decrease in the Gouy phase compared to that in standard vacuum. In the case of apertured Bessel beams, the medium allows to overcome finite aperture effects so that the corresponding diffraction length is increased; besides, the central spot of a zero order Bessel concentrates an increased fraction of the energy of the beam. Multiple scenarios beyond an electromagnetic field in the presence of an anisotropic medium could support the results reported here. They include a bosonic field in a weak gravitational field or a nontrivial spacetime background arising from Lorentz symmetry breaking. In particular, our results could illustrate how optically transparent multiferroic materials offer unprecedented opportunities to tailor structured beam propagation, as well as to simulate nontrivial spacetime backgrounds.
\end{abstract}

\maketitle

\section{Introduction}

The study of light propagating through different media is important due to its exceptional technological applications in optical communication systems as well as imaging and remote sensing. In recent years, with the advent of new methods to produce and characterize general laser beams, the study of beam propagation has taken a remarkably upgrowth. This evolution has been pushed on by the increasing compute power and the new techniques to fabricate diffractive optical elements. As a result, a plethora of new discoveries on laser beams and their propagation properties have been developed including the control of the orbital angular momentum of light \cite{PhysRevA.45.8185, Dorrah_2021, PhysRevB.88.205132}, the generation of propagation invariant beams \cite{Mazilu, PhysRevA.103.063502, PhysRevA.90.013833}, and beams that propagate along curved paths \cite{Schultheiss2010, Schultheiss2016, PhysRevD.102.084013}. 

Analogue systems are used to test theories of predicted phenomena that are hard to observe directly. For example, topological phases represent an interesting test bed for high-energy physics phenomena, in particular the electric and thermoelectric transport induced by quantum anomalies, namely, chiral, parity and gravitational anomalies. Along the same line, analogue gravity in optics has come a long way, and it drove new research directions and understandings in optics. In general, one can use transformation optics to characterize a medium that would mimic aspects of the physics in curved spacetimes. This analogy has inspired the investigation of several general relativity phenomena from the point of view of metamaterial implementations: optical analogues of black holes \cite{Narimanov2009, Sheng2013, Lee2014}, Schwarzschild spacetime \cite{Thompson2010, FERNANDEZNUNEZ2016, Bekenstein2017}, cosmic strings \cite{MACKAY2010, Bulashenko2018, HU2018} and wormholes \cite{Greenleaf2007, PratCamps2015}. In this vein, the constitutive equations formulated by Plebanski are particularly important since they relate the propagation of light in a medium with that in a gravitational field \cite{Plebanski1960}. Such optical-gravitational identification is possible since the wave equation for a given component of the electric field in a type I metamaterial \cite{Smolyaninov2011} corresponds to a (2+1)D Klein-Gordon equation for a massive scalar field, with the optical axis playing the role of the time.

The interaction between spacetime curvature and electromagnetic wave dynamics gives rise to intriguing effects like the generation of Hawking radiation \cite{Ralf2002} and the Unruh effect \cite{PhysRevD.14.870}. However, extreme conditions necessary to witness these phenomena remain inaccessible for a direct experimental investigation. Nevertheless, experimental studies evincing the impact of intrinsic and extrinsic curvature of space on the evolution of light has been carried out \cite{PhysRevLett.105.143901}. Certainly, this invite us to study, in general, the propagation of optical beams in curved spacetimes, since they can be experimentally mimicked by optical devices, and not necessarily require a truly curved spacetime. In curved spacetime, the wave equation reads $\partial _{\mu} ( \sqrt{-g} g^{\mu \nu} \partial _{\nu} \phi ) = 0$, where $\phi$ is any component of the electric or magnetic fields and $g^{\mu \nu}$ is the metric tensor. This equation not only describes the dynamics of an electromagnetic pulse in curved spacetime, but it is also able to describe its propagation (i) in  nontrivial matter or (ii) in a nontrivial spacetime background induced by spontaneous Lorentz symmetry breaking. The case (i) is achieved through Plebanski constitutive relations (i.e. the metric is a function of the permittivity, permeability and/or magnetoelectric tensor), and as we shall see below, such equation can also be obtained from Maxwell electrodynamics in anisotropic matter, being $g^{\mu \nu}$ a measure of the anisotropy. The case (ii) arises from quantum gravity candidate theories, which inevitably predict the breaking of Lorentz symmetry at very high energies through constant tensors emerging from the vacuum expectation values of some tensor operators, for which the wave equation reads $g^{\mu \nu} \partial _{\mu}  \partial _{\nu} \phi = 0$. In this sense, the metric $g^{\mu \nu}$ has constant components acting as a nontrivial spacetime background. Regardless of the origin of the nontrivial background, it can be studied by using the same mathematical techniques.

In this paper we study the propagation of structured optical beams, namely, finite apertured Gaussian and Bessel beams, in a nontrivial constant background which is characterized by a generic metric $g ^{\mu\nu}$. Splitting  the metric as the sum of two terms, $g ^{\mu\nu} = \eta ^{\mu\nu} + h ^{\mu\nu}$, where $\eta^{\mu\nu}$ is the Minkowsi flat metric and $h^{\mu\nu}$ is a constant second rank tensor, we built up the Green's function (in cylindrical coordinates) of the corresponding wave equation to study the propagation of the beams. With the help of this, we analyse the main properties of the beams as they propagate: the intensity profile, the Gouy phase, and the Rayleigh range. We find that for a constant background of the form $h ^{\mu \nu}= \xi u ^{\mu} u ^{\nu}$, where $u ^{\mu}=(u^{0},{\bf{u}})$ is a four-vector characterizing the constant background and $\xi$ is a parameter that controls the strength of the background, the beams properties strongly depends on the sign and the magnitude of the parameter $\xi$.

The outline of the present paper is the following. In Section \ref{Model1} we present the theoretical framework of electrodynamics on a constant background defined by a second-rank tensor $h ^{\mu \nu}$ and some of its main properties. Section \ref{GSection1} is devoted to the derivation of the corresponding Green's functions satisfying appropriate boundary conditions. Restricting our analysis to a tensor of the form $h ^{\mu \nu}= \xi u ^{\mu} u ^{\nu}$, where $u ^{\mu}=(u^{0},{\bf{u}})$ is a four-vector characterizing the constant background, we consider two particular cases: the timelike case, defined by $u^{0}=1$ and ${\bf{u}}={\bf{0}}$, and the radial spacelike case, defined by $u^{0}=0$ and ${\bf{u}}$ pointing radially, perpendicular to the propagation axis. In Section \ref{Results1a} we study the propagation of Gaussian and Bessel beams in a nontrivial background, mainly focusing on the impact of the parameter $\xi$ upon the main properties of the beams. We finalize in Section  \ref{Final1a}  summarizing our results and discussing possible realizations with optically transparent multiferroic metamaterials that, properly designed, could mimic a nontrivial spacetime background.

\section{Theoretical Framework}
\label{Model1}

The propagation of electromagnetic waves in curved spacetime is governed by the standard Maxwell Lagrange density
\begin{equation}
\mathcal{L}=-\frac{1}{4} g_{\mu\alpha}g_{\nu\beta} F^{\mu\nu}F^{\alpha\beta},
\end{equation}
where the metric tensor $g^{\mu\nu}$ generalizes the Minkowski flat-metric $\eta ^{\mu \nu}$ [with signature $\textrm{diag}(+,-,-,-)$], and $F ^{\mu \nu} = \partial ^{\mu} A ^{\mu} - \partial ^{\nu} A ^{\mu}$ is the electromagnetic field tensor. In curved spacetime, the metric is not a constant, but can vary in space and time. In such case, the corresponding source free Maxwell equations can be expressed as $\partial _{\mu} \mathcal{D} ^{\mu \nu} = 0$, where $\mathcal{D} _{\mu \nu} = \sqrt{-g} g _{\mu \alpha} g _{\nu \beta} F ^{\alpha \beta}$ is a tensor density of weight $+1$. If the metric has constant components, i.e. in a constant spacetime background, Maxwell equations simplify to $g ^{\mu \nu}\partial _{\nu} F _{\mu \sigma} = 0$, which in terms of the four-vector potential $A _{\mu}$, under the Lorentz gauge condition $\partial ^{\mu} A _{\mu} = 0$, lead to the well-known wave equation
\begin{equation}
g ^{\mu \nu} \partial _{\mu} \partial _{\nu} A _{\sigma} = 0 .
\end{equation}
Let us assume that we can find a coordinate system in which the metric tensor can be written in the form $g ^{\mu\nu} = \eta ^{\mu \nu} + h ^{\mu \nu}$, where $\eta ^{\mu \nu}$ is the Minkowski flat-metric and $h _{\mu \nu}$ is a general nontrivial background with constant components. Unlike a metric perturbation, here $h_{\mu \nu}$ is not necessarily small. With this choice, all the components of the four-vector potential $A^{\mu}$ satisfy the modified massless Klein-Gordon equation 
\begin{equation}
\left[ \Box +h ^{\mu \nu} \partial _{\mu} \partial _{\nu} \right] \phi(x)=0,  \label{Eqmov2}
\end{equation} 
where $\Box = \partial ^{\mu} \partial _{\mu}$ stands for the D'Alembert operator and $ \phi(x)$ is generically a component of the four-vector potential. The emergence of the $h^{\mu\nu}$-tensor can be understood from different approaches. Perhaps the most obvious is that it represents a constant background field due to gravitational effects \cite{Chang2012}, arising from the zeroth-order approximation in a curved spacetime \cite{ZHANG2017}. In a different context, the $h^{\mu\nu}$-tensor emerges as a vacuum expectation value of some tensor fields, which triggers a spontaneous breaking of Lorentz symmetry. Indeed, Eq. (\ref{Eqmov2}) arises from the scalar sector of the Standard-Model Extension \cite{Kostelecky1997, Kostelecky1998}, which is a theoretical framework conceived to describe all possible deviations from Lorentz symmetry. Candidates to quantum gravity, such as string theories \cite{Kostelecky1989} or loop quantum gravity \cite{AmelinoCamelia1998}, possess mechanisms that lead to departures from Lorentz invariance. In this context,  since the background $h ^{\mu \nu}$ does not transform as a second order tensor under \textit{active} Lorentz transformations, it specifies a privileged direction in spacetime, thus implying the breakdown of Lorentz symmetry. Since no violation of this symmetry has been detected yet, it is customary to take $\vert h ^{\mu \nu} \vert \ll 1$ in all earth-based frames. Studies on Lorentz symmetry breaking effects around a cylindrical cavity with a radial background have been recently done in Ref. \cite{Oliveira:2022ugc}.

A more realistic framework which gives rise to the field equation (\ref{Eqmov2}) is light propagation in a nontrivial optical medium, or in general, in a material media. In this case, the tensor $h ^{\mu \nu}$ contains the information regarding the optical properties of the media (i.e. permittivity, permeability and/or magnetoelectricity). In this way, we can characterize such media (so far hypothetical) by establishing its constitutive relations. To this end, we start with the modified electrodynamic Lagrange density \cite{Kostelecky2002, MartinEscobar2016}
\begin{equation}
\mathcal{L} = - \frac{1}{4} F _{\mu \nu} F ^{\mu \nu} - \frac{1}{4}(k _{F}) ^{\kappa \lambda \mu \nu} F _{\kappa \lambda} F _{\mu \nu}, \label{LAgV}
\end{equation}
whose variation with respect to the electromagnetic four-potential produces the equation of motion
\begin{equation}
\partial^\mu F_{\mu\nu}+(k_F)_{\kappa\lambda\mu\nu}\partial^{\mu}F^{\kappa\lambda}=0 . \label{FieldEqK}
\end{equation}
With the particular choice for the $(k _{F}) ^{\kappa \lambda \mu \nu}$-tensor,
\begin{equation}
(k_F)^{\mu\nu\rho\sigma}=\frac{1}{2}(\eta^{\mu\rho}h^{\nu\sigma}-\eta^{\nu\rho}h^{\mu\sigma}-\eta^{\mu\sigma}h^{\nu\rho}+\eta^{\nu\sigma}h^{\mu\rho}),
\end{equation}  
we recover the equation of motion (\ref{Eqmov2}) for all components of the four-vector potential $A _{\mu}$. Interestingly, introducing the following definitions \cite{Kostelecky2002, MartinEscobar2016}
\begin{equation}
(\kappa_{DE})^{jk}=-2(k_F)^{0j0k}, \quad\quad (\kappa_{HB})^{jk}=\frac{1}{2}\epsilon^{jpq}\epsilon^{krs}(k_F)^{pqrs},\quad\quad (\kappa_{DB})^{jk}=-(\kappa_{HE})^{kj}=\epsilon^{kpq}(k_F)^{0jpq},
\label{Def12}
\end{equation}
the field equations (\ref{FieldEqK}) can be cast in the form of the standard Maxwell equations in material media, with the modified constitutive relations
\begin{eqnarray}
\label{Rel1234}
&&{\bf{D}}=(1+\kappa_{DE}) \cdot {\bf{E}} + \kappa _{DB} \cdot {\bf{B}} ,\\ \nonumber
&&{\bf{H}}=(1+\kappa_{HB}) \cdot {\bf{B}} + \kappa_{HE} \cdot {\bf{E}} .
\end{eqnarray}
Therefore, the scalar field theory modelled by the modified Klein-Gordon equation (\ref{Eqmov2}) can be understood as arising from  an electrodynamic model for anisotropic material media.  The electrodynamic properties arising from the model (\ref{LAgV}) have been extensively studied in the literature (see Ref.  \cite{PhysRevD.66.056005} and references therein), for example, birrefringence, polarization and possible experimental tests. Interestingly, the definition of the Poynting vector and energy density remain unchanged, and since the Lagrangian (\ref{LAgV}) is gauge invariant, the relationship between the electromagnetic potentials and fields are the same as well \cite{Lehnert}.

The constitutive relations (\ref{Rel1234}) can be realized in different contexts, for example, in naturally existing anisotropic and magnetoelectric matter. Also, they can be realized under demand in artificially created composite materials which exhibit unusual properties which are not found in nature, such as metamaterials and some classes of multiferroics, which are particularly attractive since they allow to simulate curved spacetime backgrounds. An interesting particular choice for the $(k _{F}) ^{\kappa \lambda \mu \nu}$-tensor, relevant for the recently discovered topological phases of quantum matter, is $(k _{F}) ^{\kappa \lambda \mu \nu} = \theta \, \epsilon ^{\kappa \lambda \mu \nu}$, where $\theta$ is a nondynamical scalar field coupling whose origin is purely quantum-mechanical and its form depends on the symmetries of the system (i.e. it is equal to $\pi$ for time-reversal invariant topological insulators \cite{PhysRevB.78.195424}, and it becomes space and time dependent for Weyl semimetals due to the breaking of time-reversal and spatial-inversion symmetries \cite{PhysRevB.88.245107}). These phases are well described by the above constitutive relations, with however $( \kappa_{DE} ) ^{ij} = ( \kappa_{HB} ) ^{ij} = 0$ and $( \kappa _{DB} ) ^{ij} = - ( \kappa _{HE} ) ^{ij} = \theta \delta ^{ij}$. On the other hand, the relations (\ref{Rel1234}) describe also the magnetoelectric response of multiferroic materials, such as Cr$_2$O$_3$, in which case the matrices $\kappa_{DE}$ and $\kappa_{HB}$ contribute to the conventional permittivity $\hat{\epsilon} (\omega )$ and permeability $\hat{\mu} (\omega )$ tensors, and the magnetoelectric tensor $\hat{\chi} ^{\mbox{\scriptsize me}} (\omega ) = \kappa _{DB} = - \kappa _{HE}$ is not topological in nature as in topological phases, but arises from the coupling between spins and electric dipoles in the material  \cite{HEHL20081141}.  All of these matrices are determined by the lattice structure of the material and they are frequency-dependent. The case of multiferroic materials could be specially relevant for the results reported in this paper, since they can be optically transparent and  act like a nontrivial spacetime background. It is to be noted that our theoretical analysis below can be applied to any of the aforementioned realizations of the constitutive relations (\ref{Rel1234}), with the appropriate choice of the matrices appearing there.

In this paper we shall consider a particular choice for the $h^{\mu\nu}$-tensor, namely
\begin{equation}
h ^{\mu \nu} = \xi u ^{\mu} u ^{\nu} , \label{back122}
\end{equation} 
where $\xi$ is a parameter that controls the strength of the nontrivial background, and $u ^{\mu}=(u^{0},{\bf{u}})$ is a four-vector that specifies its direction. The definitions in Eq. (\ref{Def12}), which establish the constitutive relations in Eq. (\ref{Rel1234}), provide us the relations
\begin{equation}
(\kappa_{DE}) ^{jk} \rightarrow - \xi (u ^{j} u ^{k} + u _{0} ^{2} \, \eta ^{jk}), \qquad ( \kappa _{HB}) ^{jk} \rightarrow\xi (u_0^2-u_\mu u^\mu)\eta^{jk}+\xi u^j u^k, \qquad (\kappa_{DB})^{jk}=-(\kappa_{HE})^{kj}\rightarrow \xi\, u_0u_i\epsilon^{ijk} .
\end{equation}
With the help of Eq. (\ref{back122}), the equation of motion (\ref{Eqmov2}) simplifies to
\begin{align}
\left[ \Box + \xi \left( u _{\mu} \partial ^{\mu} \right) ^{2} \right] \phi (x)  = 0 .
\label{EQLV}
\end{align}
This theory has been explored extensively as a Lorentz-violating extension of the scalar field theory. For example, it has been used to compute the Casimir pressure between parallel conducting plates embedded in a nontrivial background modelled by the particular tensor (\ref{back122}) \cite{PhysRevD.96.045019, doi:10.1142/S0217732318501158, PhysRevD.101.095011, PhysRevD.95.025021}, as well as in spherical \cite{PhysRevD.102.015027} and cylindrical \cite{doi:10.1142/S0217751X21501682} configurations. Also, the Casimir effect between parallel plates with a general symmetric-background tensor $h ^{\mu \nu}$ has been computed in Ref. \cite{ESCOBAR2020135567}.

Therefore, regardless of the origin of the tensor $h ^{\mu \nu}$, it is convenient to work in an axiallly symmetric spacetime, such that the principal axis coincides with the direction of the optical beam propagation, i.e. ${\bf{k}} \cdot {\bf{u}}=0$, where ${\bf{k}}$ is the wave-vector of the beam and ${\bf{u}}$ as defined in Eq. (\ref{back122}) specifies a preferred spatial direction. Cylindrical spacetimes has been extensively studied in general relativity \cite{Bronnikov2020}. For example, the Levi Civita \cite{LeviCivita2011} and Lewis \cite{Lewis1932} spacetimes represent vacuum solutions to the Einstein field equations corresponding to a static cylindrical vacuum spacetime and a stationary cylindrical vacuum spacetime, respectively. Sources of the later are discussed in Refs. \cite{Pereira2000, Silva2002}. Topics as gravitational waves from rotating cylindrical sources \cite{Mashhoon2000}, the gravitational collapse in cylindrical symmetry \cite{Mena2015} and cylindrical wormholes \cite{Bronnikov2016, Bronnikov2016.2}  have been also studied. Even more, cylindrical symmetry has been used to model, as a first order approximation, the rotating matter at the center of galaxies \cite{Pereira2000}. 

From the condensed matter side, the optical response of materials with cylindrical symmetry obeying the above constitutive relations, has been widely explored.  For example,  the magnetoelectric effect in cylindrical topological insulators is investigated in Ref. \cite{PhysRevD.98.056012}. Also, the boundary value problem of a multiferroic composite cylinder structure consisting of bonded piezoelectric/piezomagnetic cylinders is considered in Ref. \cite{doi:10.1080/15397734.2021.1929312}, where the direct magnetoelectric effect is investigated.  In view of the above, the mathematical tools required to analyze the propagation of light in both nontrivial axially-symmetric background and material media with cylindrical geometry, are quite similar, and the difference lies only in the physical origin of the tensor $h^{\mu \nu}$.

For definiteness,  here we shall consider the four-vector $u ^{\mu}$ in Eq. (\ref{back122}) possessing axial symmetry around the direction of beam propagation, that we take along the $z$-direction. To be precise, we take two particular cases: (i) the timelike case for which $u _{0}=1$ and ${\bf{u}} \neq {\bf{0}}$, and the radial spacelike case for which $u _{0}=0$ and ${\bf{u}}$ pointing along the radial direction.

For monochromatic waves,  we look for solutions of Eq. (\ref{EQLV}) of the form $\phi ({\bf{r}},t) = \int d \omega \, \phi _{\omega} ({\bf{r}}) e ^{- i \omega t}$, where $\omega$ is the angular frequency and $\phi _{\omega} ({\bf{r}})$ is the frequency-dependent complex amplitude of the field. The substitution of this solution into Eq. (\ref{EQLV})  leads to the modified Helmholtz equation
\begin{align}
\left\lbrace ( \omega / c ) ^{2} + \nabla ^{2} + \xi \left[ - i u _{0}  ( \omega / c ) - {\bf{u}} \cdot \nabla \right] ^{2} \right\rbrace \phi _{\omega} ({\bf{r}})  = 0.
\label{Helmholtz}
\end{align}
Therefore, the complex amplitude at propagation distance $z$ from the source at $z ^{\prime}$ is given by the integral \cite{MartinD}:
\begin{align}
 \phi _{\omega} ( {\bf{r}} _{\perp},z ) = \int G _{\omega}  ({\bf{r}} _{\perp} , z ; {\bf{r}} _{\perp} ^{\, \prime} , z ^{\prime} ) \,  \phi _{\omega}  ({\bf{r}} _{\perp} ^{\, \prime} , z ^{\, \prime} ) \, d ^{2} {\bf{r}} ^{\, \prime} _{\perp},
\label{Prop}
\end{align}
where ${\bf{r}} _{\perp} = (x,y)$ is the transverse position to the propagation axis, $\phi _{\omega} ({\bf{r}} _{\perp} ^{\, \prime} , z ^{\, \prime} ) $ is the optical wave at the source and $G _{\omega} ({\bf{r}} _{\perp} , z ; {\bf{r}} _{\perp} ^{\, \prime} , z ^{\prime} )$ is the corresponding Green's function. In the next section we will compute the Green's functions that allow us to investigate the influence of the nontrivial background on the propagation of optical beams.

The study of light propagation in general requires vector solutions to Helmholtz equation, thus allowing the identification of the polarization states.  For example, the propagation of vector Bessel beam  \cite{Volke_Sepulveda_2006} and vector Helmholtz-Gauss beams \cite{Bandres:05} have been analysed in the Lorentz-symmetric case.  It would be of wide interest to include the Lorentz-violating effects in the propagation of vector beams. Here, we just consider the propagation of scalar beams, which is appropriate for linearly polarized light under certain circumstances \cite{PhysRevA.11.1365}.

\section{Green's Function}
\label{GSection1}

In this section we derive the Green's function $ G _{\omega} ({\bf{r}} _{\perp} , z ; {\bf{r}} _{\perp} ^{\, \prime} , z ^{\prime} )$ appearing in Eq. (\ref{Prop}) for the radial spacelike and timelike cases.

 
\subsection{Radial spacelike case}

We first analyse the radial spacelike case, which is defined by the constant background vector pointing along the radial space direction. According to the axial symmetry of the problem, it is appropriate to evaluate the Green's function in the cylindrical coordinates $(\rho,\theta,z)$, where  $z$ is the propagation axis and $\rho = \vert {\bf{r}} _{\perp}  \vert $ is the transverse position.  In the cylindrical coordinates,  Eq. (\ref{EQLV}) implies that the Green's function should satisfy 
\begin{align}
\left[ ( \omega / c ) ^{2} + \frac{\partial^2}{\partial z^2} + (1-\xi)\frac{\partial ^2}{\partial \rho^2}+\frac{1}{\rho}\frac{\partial}{\partial \rho}+\frac{1}{\rho^2}\frac{\partial^2}{\partial\theta^2} \right] G _{\omega} ({\bf{r}} _{\perp} , z ; {\bf{r}} _{\perp} ^{\, \prime} , z ^{\prime} )  = - \frac{1}{\rho } \delta (\rho  - \rho ^{\prime}) \delta (z-z ^{\prime})\, \delta (\theta - \theta ^{\prime})  .
\end{align}
Owing to the axial symmetry and the translational invariance along the $z$-axis, we can introduce suitable representations for the Dirac delta functions $\delta (z-z ^{\prime})$ and $\delta (\theta - \theta ^{\prime})$, namely,
\begin{align}
\delta (z-z ^{\prime}) = \int _{- \infty} ^{\infty} \frac{dk}{2 \pi} e ^{i k (z-z ^{\prime})}  , \qquad \delta (\theta - \theta ^{\prime}) = \frac{1}{2 \pi} \sum _{m = - \infty} ^{\infty} e ^{im (\theta - \theta ^{\prime})} . \label{DeltaRepresentations}
\end{align}
Exploiting these symmetries we further introduce a Fourier representation of the Green's function as
\begin{align}
G _{\omega} ({\bf{r}} _{\perp} , z ; {\bf{r}} _{\perp} ^{\, \prime} , z ^{\prime} ) =  \sum _{m = - \infty} ^{\infty} \int _{- \infty} ^{\infty} \frac{dk}{(2 \pi) ^{2}} \, e ^{i k (z-z ^{\prime})} e ^{im (\theta - \theta    ^{\prime})} \, g _{m} (\rho , \rho ^{\prime} ; k , \omega ), \label{GreenExpansion}
\end{align}
where $g _{m} (\rho , \rho ^{\prime} ; k  , \omega )$ is the reduced Green's function. With all the above, it is found that the reduced GF satisfies the differential equation 
\begin{align}
\left[ ( \omega / c ) ^{2} - k ^{2} + (1-\xi)\frac{\partial ^2}{\partial \rho ^{2}}+\frac{1}{\rho}\frac{\partial}{\partial \rho} - \frac{m ^{2}}{\rho ^{2}} \right] g _{m} (\rho , \rho ^{\prime} ; k  , \omega ) = - \frac{ \delta (\rho  - \rho ^{\prime})}{\rho} . \label{Eqgm}
\end{align}
This equation has to be solved subject to the physical boundary conditions that $g _{m}$ be bounded both at the origin and as $\rho$ recedes to infinity. The resulting equation (\ref{Eqgm}) 
acquires a precise meaning when converting it into a boundary value problem.  Heuristically, this equation can be interpreted as consisting of the homogeneous differential equation (defined for $\rho \in \mathbb{R} ^{+} \setminus \{\ \!\! \rho ^{\prime} \}\ $)
\begin{align}
\left[ ( \omega / c ) ^{2} - k ^{2}  + (1 - \xi ) \frac{\partial ^{2}}{\partial \rho ^{2}}+\frac{1}{\rho} \frac{\partial}{\partial \rho} - \frac{m ^{2}}{\rho ^{2}} \right] g _{m} (\rho , \rho ^{\prime} ; k  , \omega  )  = 0
\label{B2N}
\end{align}
together with boundary conditions for $g _{m}$ and its derivative $\partial _{\rho} g _{m}$ at the singular point $\rho = \rho ^{\prime}$. If we merely accept that $g _{m}$ is bounded when $\rho$ is in the infinitesimal neighbourhood of $\rho ^{\prime}$, integration of Eq. (\ref{Eqgm}) over the interval between $\rho ^{\prime} - \epsilon$ and $\rho ^{\prime} + \epsilon$, with $\epsilon \to 0 ^{+}$, yields
\begin{align}
\frac{\partial g _{m} (\rho , \rho ^{\prime} ; k  , \omega  )}{\partial \rho} \bigg| ^{\rho = \rho ^{\prime} + \epsilon}  _{\rho = \rho ^{\prime} - \epsilon} &= - \frac{1}{\rho ^{\prime}} \, \frac{1}{1 - \xi} .
\label{dgdisc}
\end{align}
\begin{figure}
\centering
\includegraphics[width=0.45\textwidth]{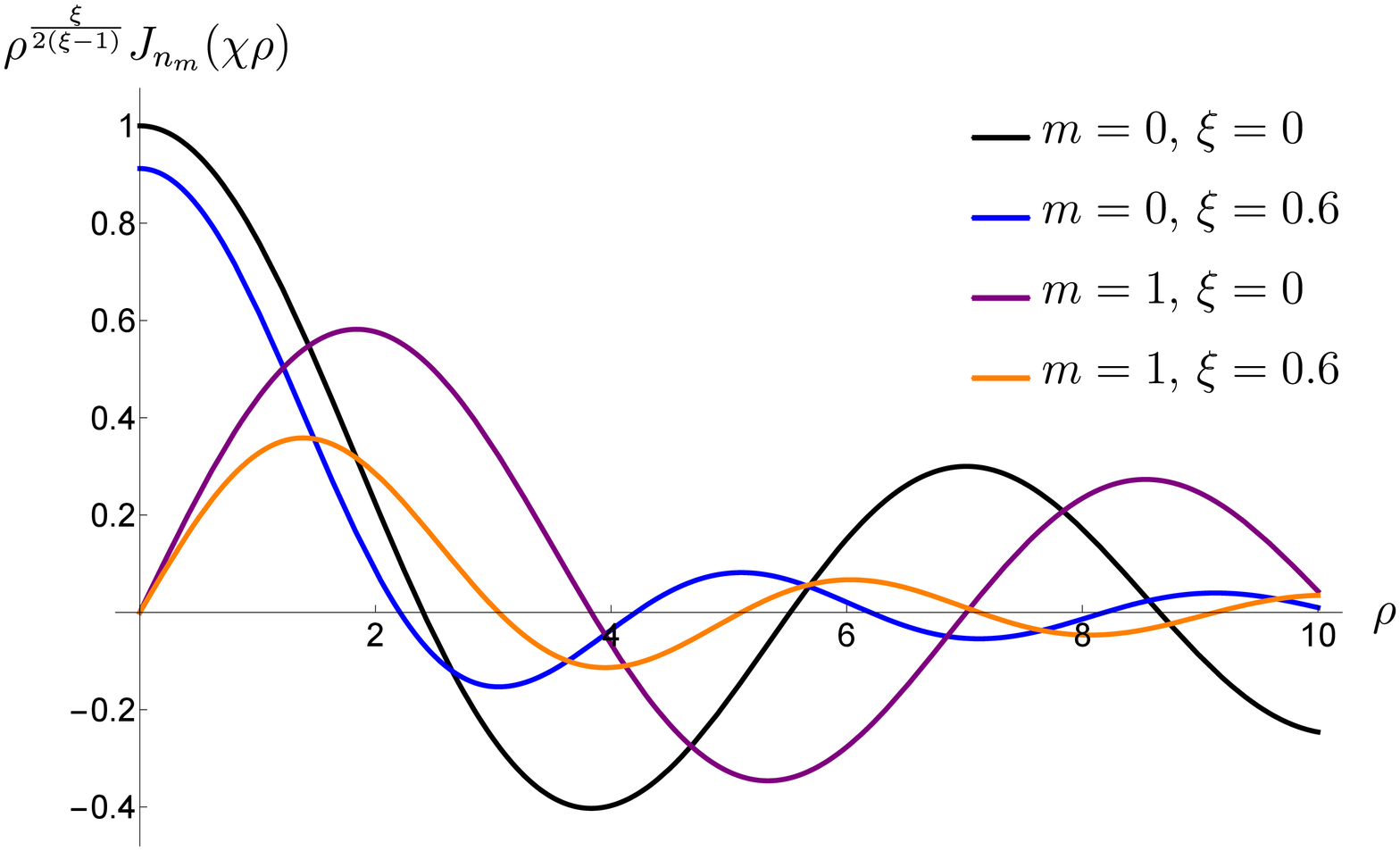} \qquad
\includegraphics[width=0.45\textwidth]{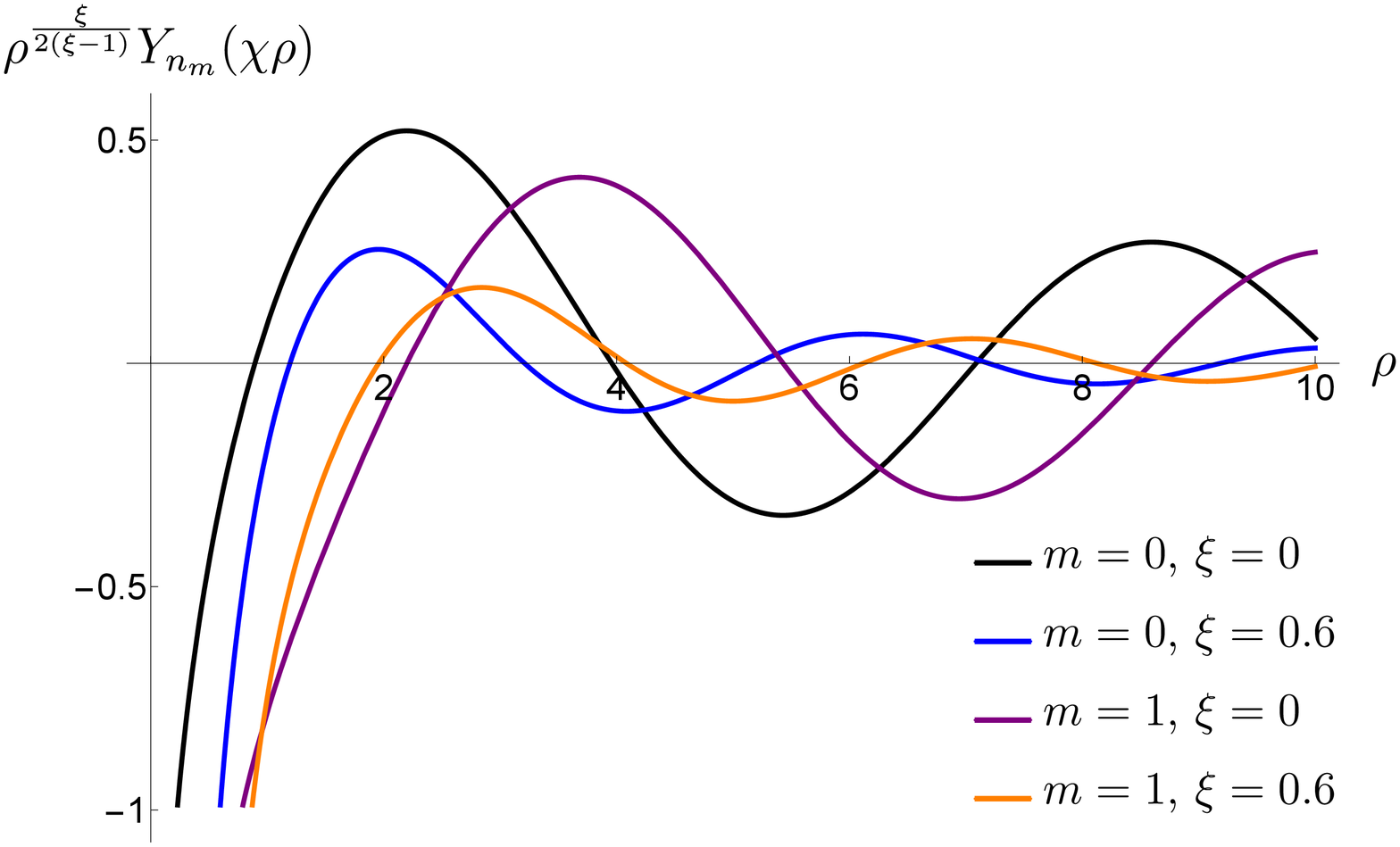}
\caption{The two linear independent solutions to the homogeneous equation (\ref{B2N}) with $q=1$ can be written in terms of standard Bessel functions multiplied by a factor that depends on the metric parameter $\xi$. This parameter also determines the effective dispersion relation and the Bessel functions order.}
\label{Fig1}
\end{figure} 
Then, the continuity of $g _{m}$ at $\rho = \rho ^{\prime}$ follows. The solution to the homogeneous equation (\ref{B2N}) can be written in terms of the Bessel functions of noninteger order:
\begin{align}
    g _{m} (\rho , \rho ^{\prime} ; k , \omega ) =  \rho ^{\frac{\xi}{2(\xi - 1)}} \, \left[ c _{1} \, J _{n _{m}} (  \chi \rho  ) + c _{2} \, Y _{n _{m}} (  \chi \rho  ) \right] ,
    \label{gmH}
\end{align}
where
\begin{align}
\chi = \sqrt{\frac{ ( \omega / c ) ^{2} - k ^{2}}{1 - \xi}} , \qquad n _{m} = \frac{\sqrt{4m ^{2} (1 - \xi) + \xi ^{2}}}{2 ( 1 - \xi)} ,  \label{def1}
\end{align}
which correctly reduces to the standard results in the limit $\xi \to 0 $, i.e. $n _{m} = m$ and $\chi = \sqrt{ ( \omega / c ) ^{2} - k ^{2}}$. The choice between the two independent solutions appearing in Eq. (\ref{gmH}) is dictated by the physical requirements at the origin and at the infinity. In Fig. \ref{Fig1} we show the two independent solutions $ \rho ^{\frac{\xi}{2(\xi - 1)}} \, J _{n _{m}} (  \chi \rho  )$ and  $ \rho ^{\frac{\xi}{2(\xi - 1)}} \, Y _{n _{m}} (  \chi \rho  )$, as a function of $\rho$ and for different values of the parameter $\xi$. These solutions preserve their oscillatory and asymptotic behaviour. As a consequence of this, now we can solve for the reduced GF. Finiteness of the GF at the origin, as well as the outgoing wave solution at infinity, imply that it can be taken as
\begin{align}
g _{m} (\rho , \rho ^{\prime} ; k  , \omega ) = \rho ^{\frac{\xi}{2(\xi - 1)}} \left\lbrace \begin{array}{l} A _{m} ( \rho ^{\prime} ) \, J _{n _{m}} ( \chi \rho  ) \\[7pt] B _{m} ( \rho ^{\prime} ) \, H _{n _{m}} ^{(1)} ( \chi \rho  ) \end{array} \begin{array}{l} \mbox{for } \rho < \rho ^{\prime} \\[7pt]  \mbox{for } \rho > \rho ^{\prime} \end{array} \right. ,
\end{align}
where $H _{n _{m}} ^{(1)} $ is the Hankel function of the first kind. The coefficients $A _{m}$ and $B _{m}$ have to be determined by imposing the boundary conditions at the singular point $\rho = \rho ^{\prime}$. The final result is
\begin{align}
g _{m} (\rho , \rho ^{\prime} ; k  , \omega ) = \frac{i ( \pi / 2)}{1 - \xi} \, \left( \frac{\rho}{\rho ^{\prime}} \right) ^{\frac{\xi}{2(\xi - 1)}}  J _{n _{m}} ( \chi \rho _{<}  ) \, H _{n _{m}} ^{(1)} ( \chi \rho _{>} ),
\label{gmR}
\end{align}
where $\rho_>$ ($\rho_<$) is the greater (lesser) between $\rho$ and $\rho ^{\prime}$. In Fig. \ref{Figx} we plot the reduced GF of Eq. (\ref{gmR}) as a function of $\rho$ and different values of the parameter $\xi$. As we can see, the reduced GF has the expected behaviour at the singular point as well as at the origin and at infinity. All in all, we obtain for the Green's function in the full coordinate space
\begin{align}
G _{\omega} ({\bf{r}} _{\perp} , z ; {\bf{r}} _{\perp} ^{\prime} , z ^{\prime} )  = \frac{i ( \pi / 2)}{1 - \xi}  \left( \frac{\rho}{\rho ^{\prime}} \right) ^{\frac{\xi}{2(\xi - 1)}} \sum _{m = - \infty} ^{\infty} \int _{- \infty} ^{\infty} \frac{dk}{(2 \pi) ^{2}}  \, e ^{i k (z-z ^{\prime})} e ^{im (\theta - \theta    ^{\prime})} J _{n _{m}} ( \chi \rho _{<}  ) \, H _{n _{m}} ^{(1)} ( \chi \rho _{>} ) .
  \label{GreenFR}
\end{align}
With the Green's function (\ref{GreenFR}) we are able to analyse the propagation of optical beams in a nontrivial background for any initial input optical wave. 
\begin{figure}
\centering
\includegraphics[width=0.4\textwidth]{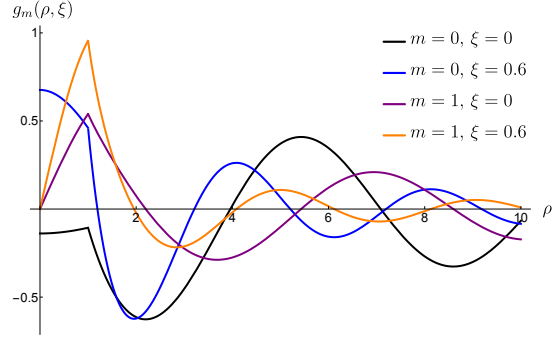}
\caption{The reduced Green functions $g_m$ in Eq. (\ref{gmR}) with $q=1$ and $\rho^\prime=1$}
\label{Figx}
\end{figure} 

\subsection{Timelike case}

We now turn to the timelike case, which is defined by the background fourvector $u ^{\mu} = (1,0,0,0)$. Then Eq. (\ref{Helmholtz}) becomes
\begin{align}
\left[ (1 + \xi ) ( \omega / c ) ^{2} + \frac{\partial ^{2}}{\partial \rho ^{2}} + \frac{1}{\rho} \frac{\partial}{\partial \rho} + \frac{1}{\rho ^{2}} \frac{\partial ^{2}}{\partial \theta ^{2}} + \frac{\partial ^{2}}{\partial z ^{2}} \right] \phi _{\omega} ({\bf{r}}) = 0 .
\end{align}
We observe that, unlike the radial spacelike case where the background contribution, encoded in $\xi$, enters into the Green's function (\ref{GreenFR}) in a nontrivial way, in the timelike case the $\xi$-dependence can be absorbed into the frequency through the redefinition $\omega \sqrt{ 1 + \xi } \to \omega$, thus leaving us with the standard Green's function equation.  Physically, this can be interpreted as a modification in the speed of light, which potentially can be tested in experiments. However, from the mathematical point of view, the beam propagation is carried out such as it happens in a trivial vacuum spacetime.  The consequences of this selection are interesting though evident so that they do not require at this stage further theoretical study. Therefore, hereafter we will primarily focus on the radial spacelike case.

\section{Optical beam propagation}
\label{Results1a}

In the previous section we have derived the propagator or Green's function which will allow us to model light propagation in the presence of a nontrivial background.  The quantitative study is carried out by evaluating numerically the intensity of the optical wave $I({\bf{r}} _{\perp} ,z) = \vert \phi _{\omega} ( {\bf{r}} _{\perp},z ) \vert ^{2} $ on the basis of the analytic formula (\ref{Prop}) with the propagator derived above. Because we assume no power loss, the total power $P (z) = \int d ^{2} {\bf{r}} _{\perp} \, I({\bf{r}} _{\perp} ,z) $ will be preserved during propagation, independently of the parameter $\xi$.  In what follows, we consider two particular examples of beam propagation, namely, finite apertured Gaussian and Bessel beams.

\subsection{Gaussian beam}

The Gaussian beam (GB) is a fundamental model in optics with a wide variety of applications. It represents a good approximation to the electric field amplitude of the fundamental mode at the output of a standard laser \cite{Siegman}, as well as to the electric amplitude in weakly guiding waveguides with cylindrical symmetry  \cite{Constantinou}. In practice, it is used in innumerable experimental devices,  within the scope of this work we can mention laser stabilization cavities \cite{DavilaRodriguez2017} and gravitational wave detectors \cite{Saulson}.  In fact, Gaussian beams have been considered as a test bed for the impact of Earth's gravity on beam propagation \cite{Ulbricht2020, Ulbricht2021}.


A GB can be described as a scalar wave whose wave fronts are predominantly transverse to the main direction of propagation, here taken as the $z$-axis, with the perpendicular components of the wave vector $\mathbf{k}_\bot$ exhibiting a Gaussian distribution. In most actual implementations $\vert \mathbf{k}_\bot\vert \ll \vert k_z\vert$ so that the paraxial approximation is adequate. 
The amplitude distribution of an apertured Gaussian beam of waist size $w _{0}$ at the transmitting aperture located at $z = 0$  is taken as \cite{Yariv}
\begin{equation}
\phi _{\omega} ( {\bf{r}} _{\perp} , 0 ) =  \sqrt{\frac{2 I _{0}}{\pi w _{0} ^{2}}} 
\frac{1}{\sqrt{1 - \exp \left( - \frac{2\rho _{0} ^{2}}{w _{0} ^{2}} \right) } } \, \exp \left( - \frac{\rho ^{2}}{w _{0} ^{2}} \right) \, \Theta( \rho _{0} - \rho ) ,
\label{hazgaussiano}
\end{equation}
where $\rho = \sqrt{x ^{2} + y ^{2}}$ is the radial distance from the beam center line and $\Theta (x)$ is the Heaviside function which expresses the condition for a finite circular aperture of radius $\rho _{0}$. The input beam (\ref{hazgaussiano}) is properly normalized to $I _{0}$, i.e. $\int d ^{2} {\bf{r}} _{\perp} \, \vert \phi _{\omega} ( {\bf{r}} _{\perp} , 0 ) \vert ^{2} =  I _{0}$, where $I _{0}$ is the power of the beam.

\begin{figure}
\subfloat[\label{ac1}$\xi=0$]{\includegraphics[width=0.33\textwidth]{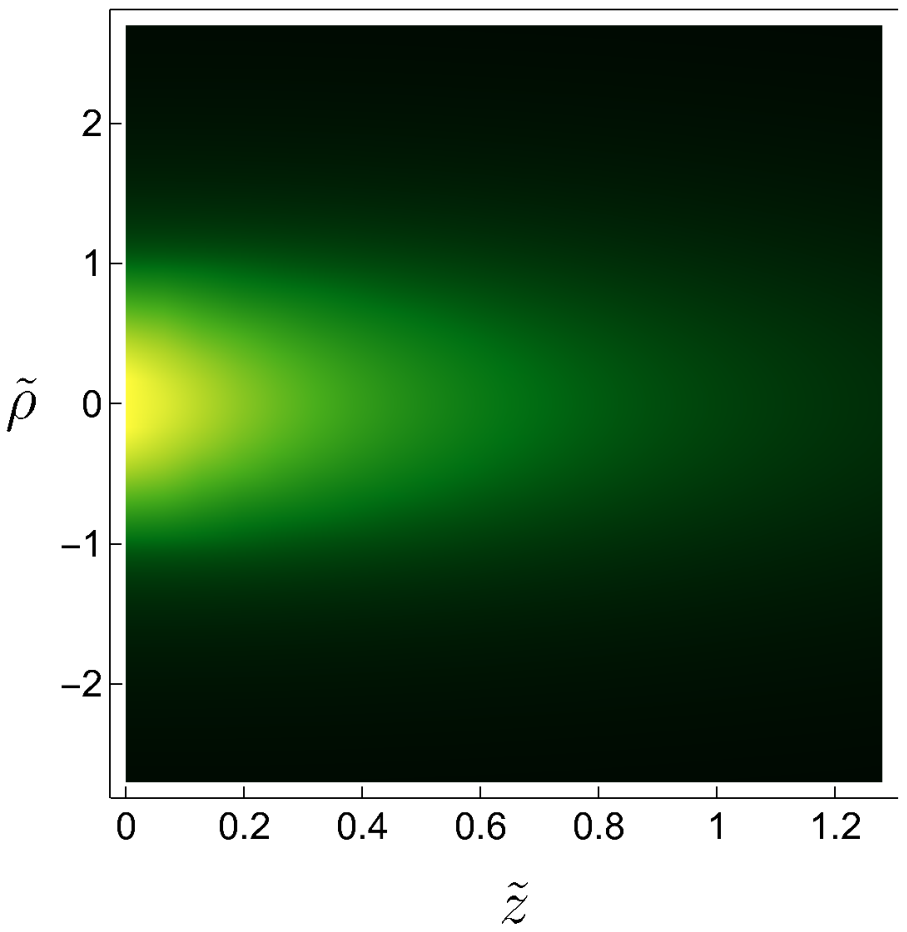}} \hspace{1cm}
\subfloat[\label{ac2}$\xi=0.4$]{\includegraphics[width=0.39\textwidth]{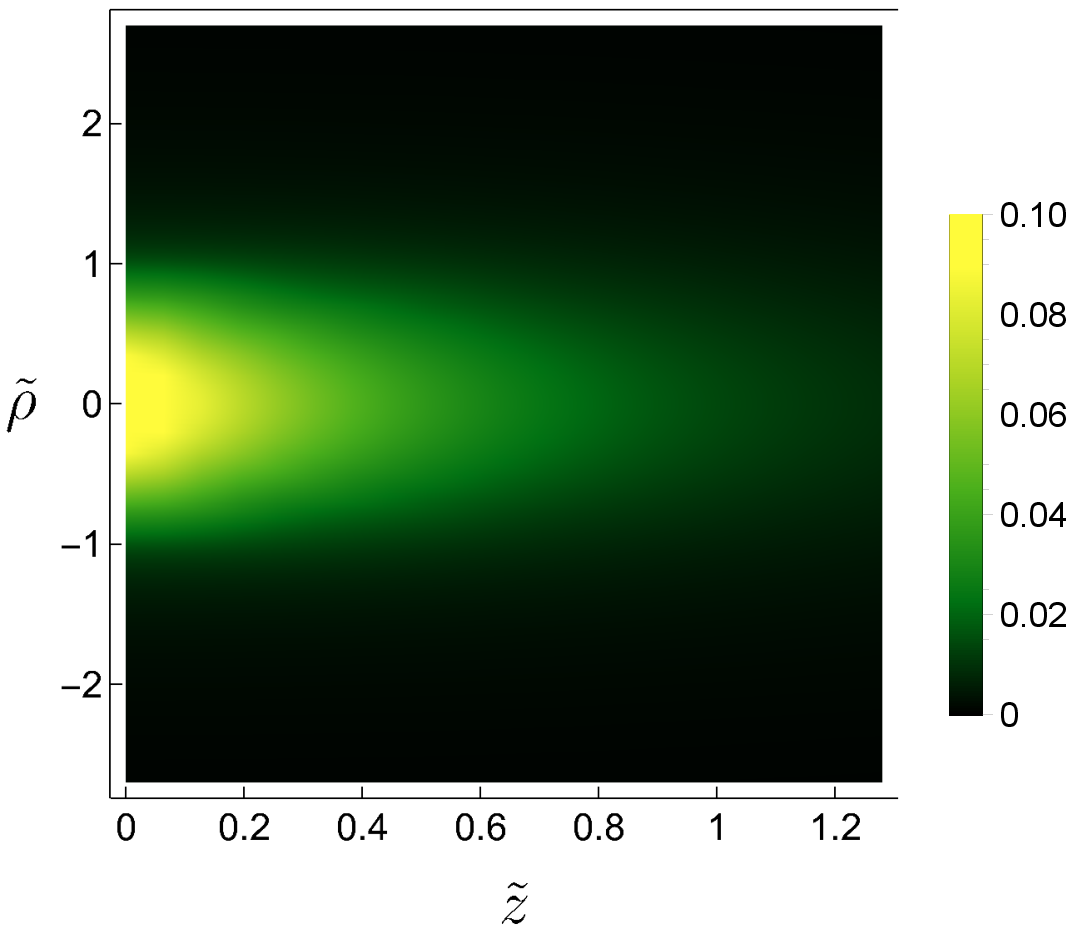}}
\caption{Intensity of the field $I ( \tilde{\rho}, \tilde{z} )$ (in units of $ I _{0} $) as a function of the dimensionless coordinates $\tilde{\rho}$ and $\tilde{z}$ for $\xi = 0$ (in vacuum) and $\xi = 0.4$ (a highly exagerated nontrivial background parameter).} \label{DensityG1}
\end{figure}

\begin{figure}
\centering
\includegraphics[width=0.4\textwidth]{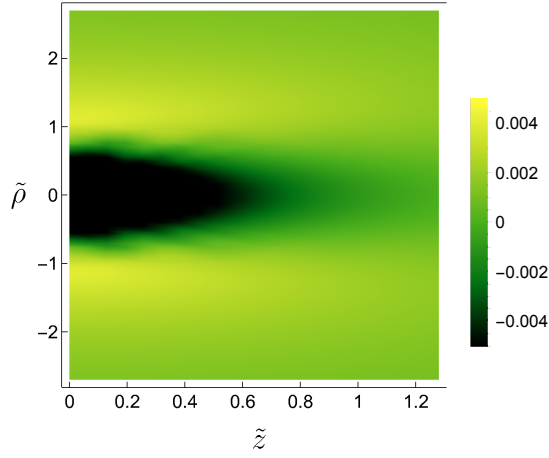}
\caption{Difference between the intensities $I( \tilde{\rho}, \tilde{z} )_{\xi = 0}-I( \tilde{\rho}, \tilde{z} )_{\xi = 0.4}$ (in units of $ I _{0} $) as a function of the dimensionless coordinates $\tilde{\rho}$ and $\tilde{z}$.}
\label{DensityG1ab}
\end{figure}

\begin{figure}
\centering
\subfloat[\label{dc1}$\tilde{z}=0$]{\includegraphics[width=0.25 \textwidth]{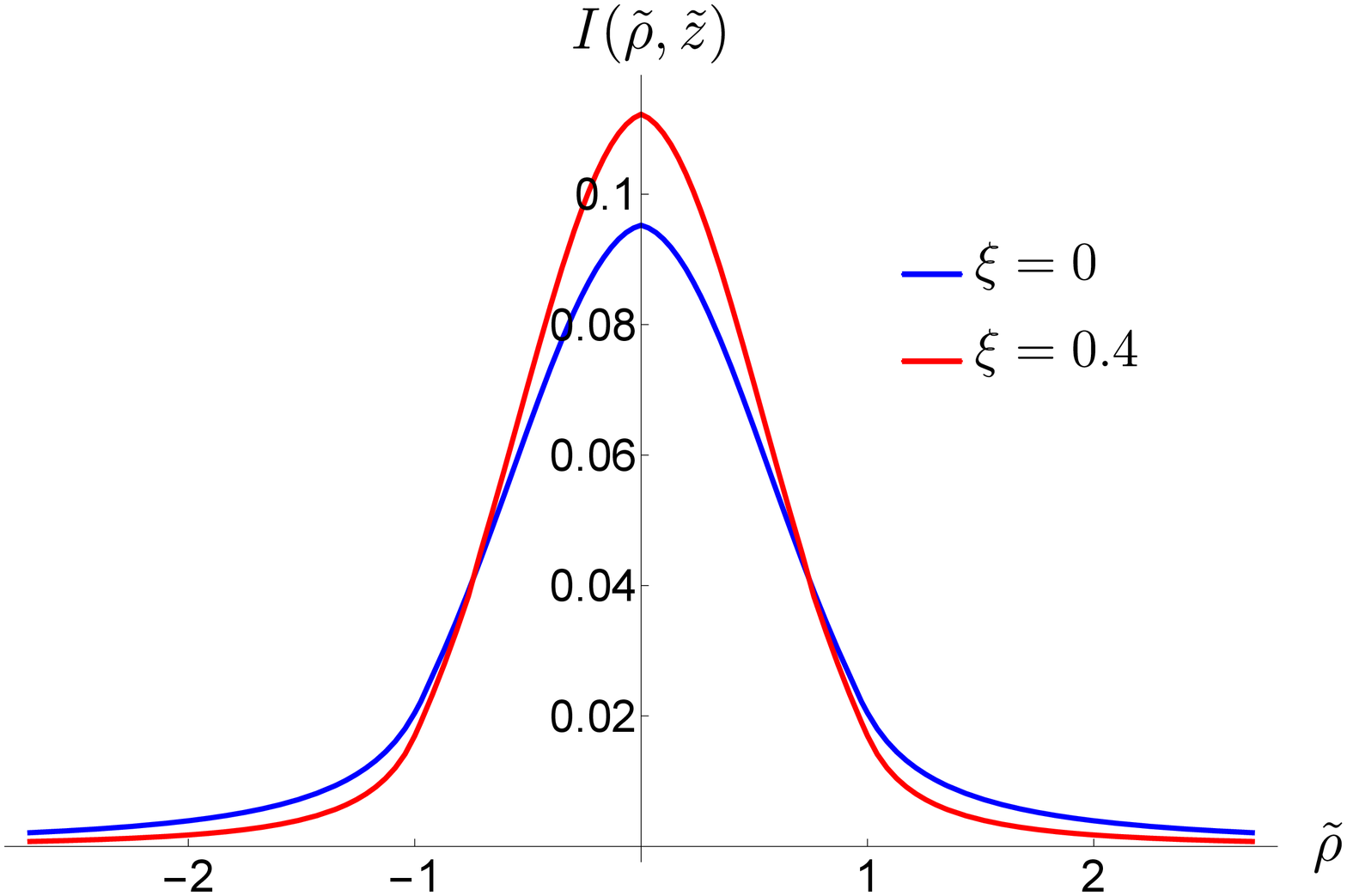}} \hfill
\subfloat[\label{dc2}$\tilde{z}=0.384$]{\includegraphics[width=0.25 \textwidth]{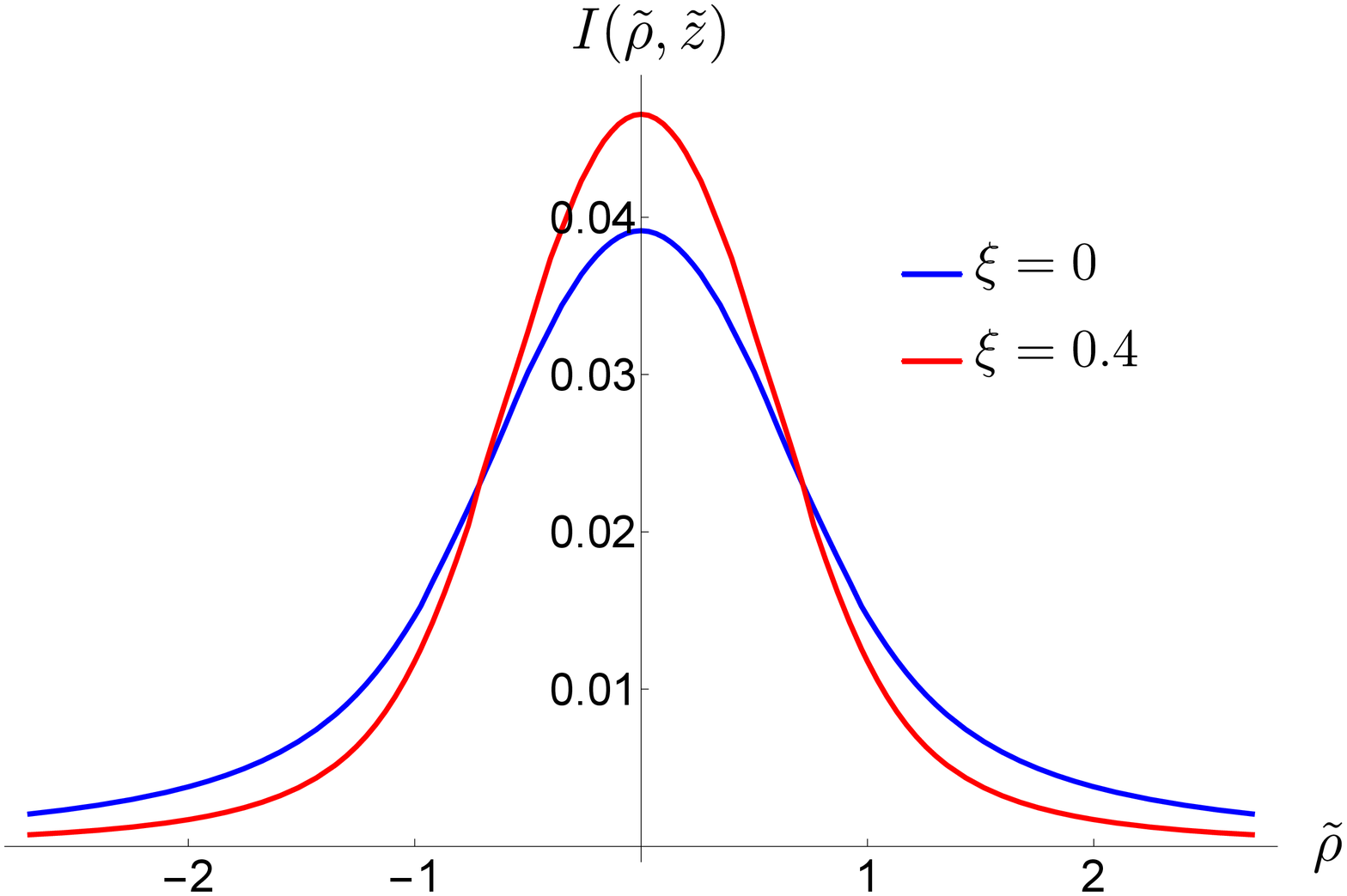}} \hfill
\subfloat[\label{dc3}$\tilde{z}=0.768$]{\includegraphics[width=0.25 \textwidth]{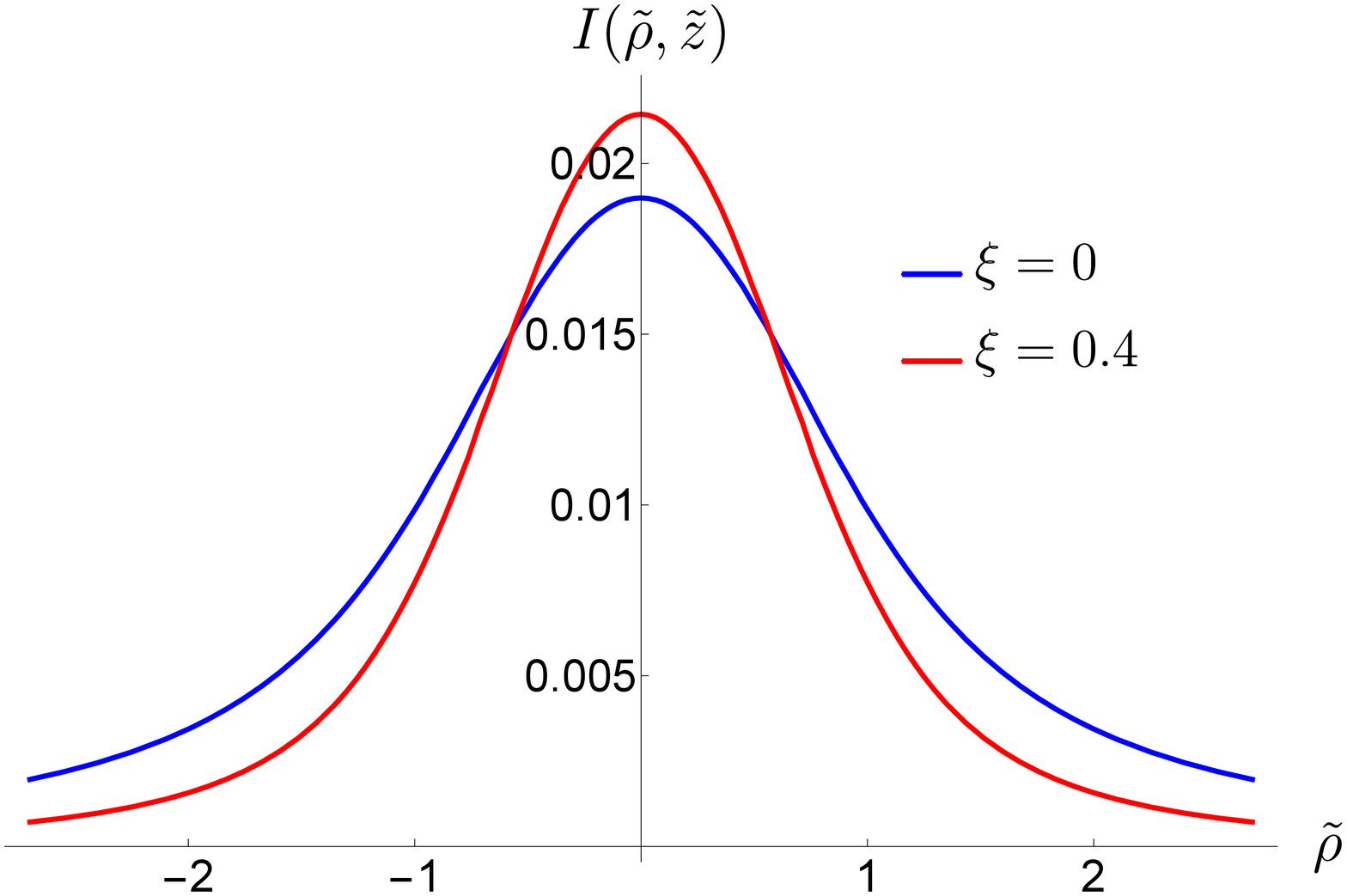}} \hfill
\subfloat[\label{dc4}$\tilde{z}=1.088$]{\includegraphics[width=0.25 \textwidth]{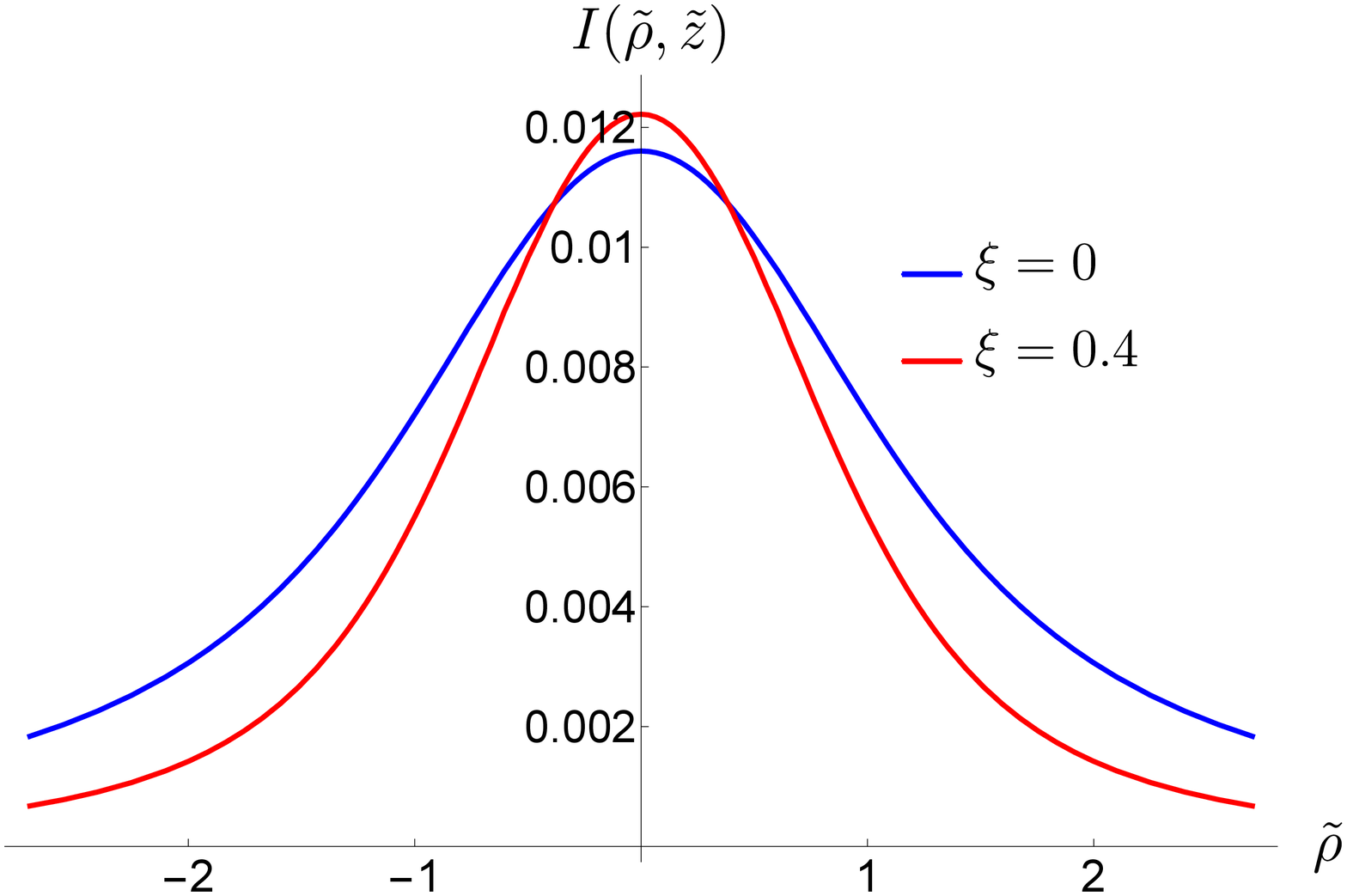}}\
\caption{Intensity profile $I ( \tilde{\rho}, \tilde{z} )$ (in units of $ I _{0} $) as a function of the dimensionless radial coordinate $\tilde{\rho}$, for $\xi = 0$ and $\xi = 0.4$, and different values of the dimensionless propagation distance $\tilde{z}$.} \label{Fig3m}
\end{figure}

\begin{figure}[h]
\centering
\subfloat[\label{EjeRho0}]{\includegraphics[width=0.33 \textwidth]{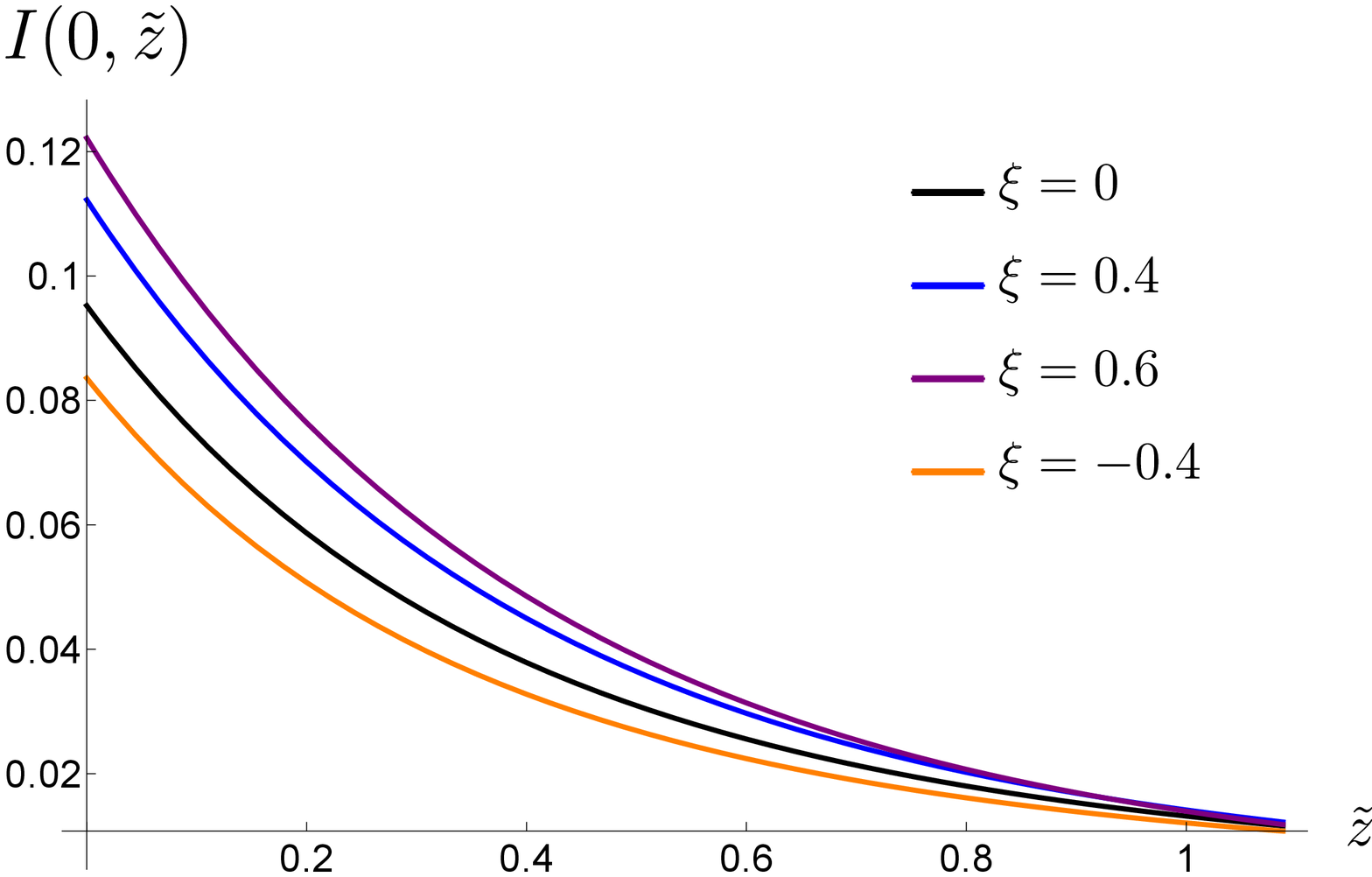}} \quad
\subfloat[\label{ZRay1}]{\includegraphics[width=0.29 \textwidth]{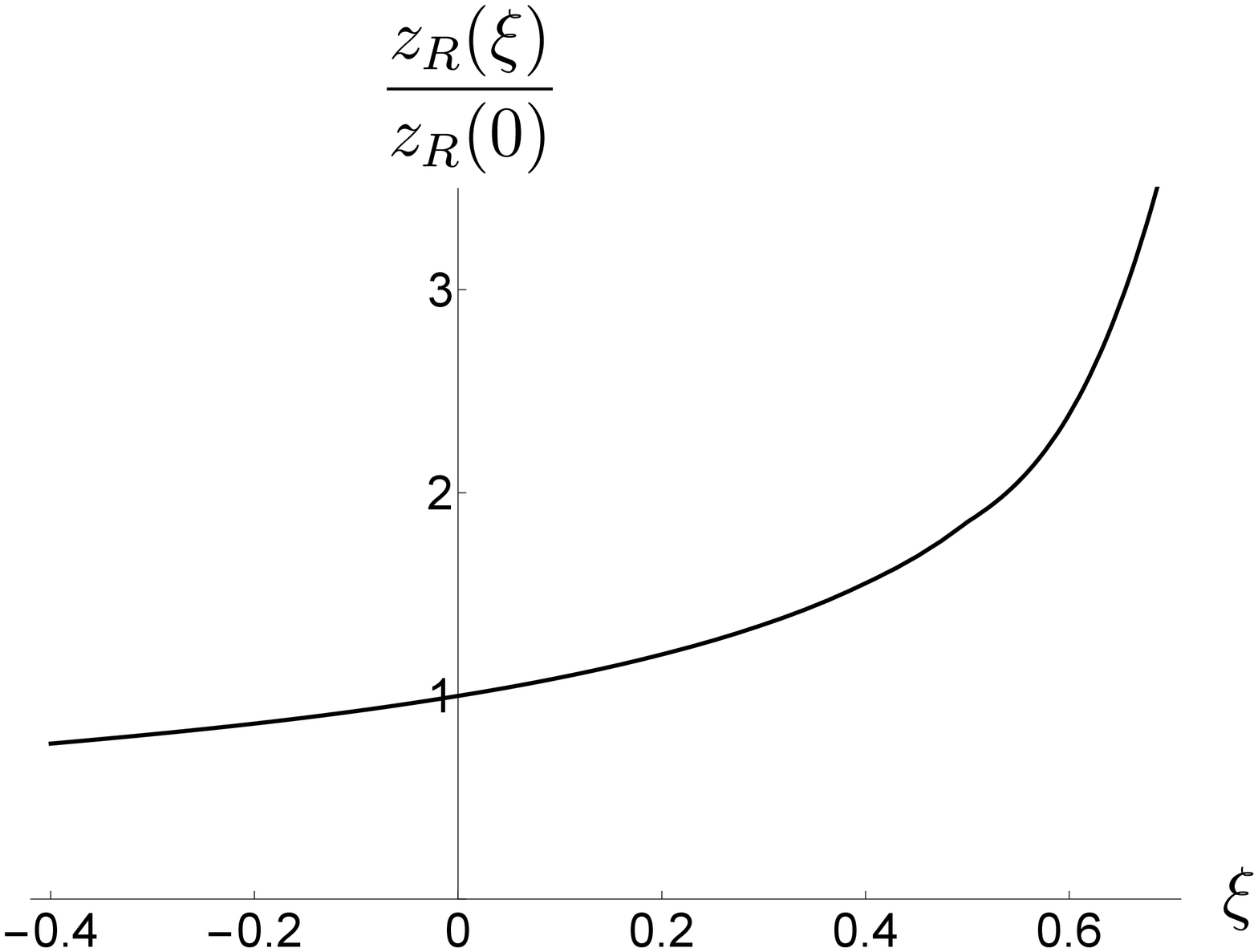}} \quad
\subfloat[\label{Gouy123}]{\includegraphics[width=0.33 \textwidth]{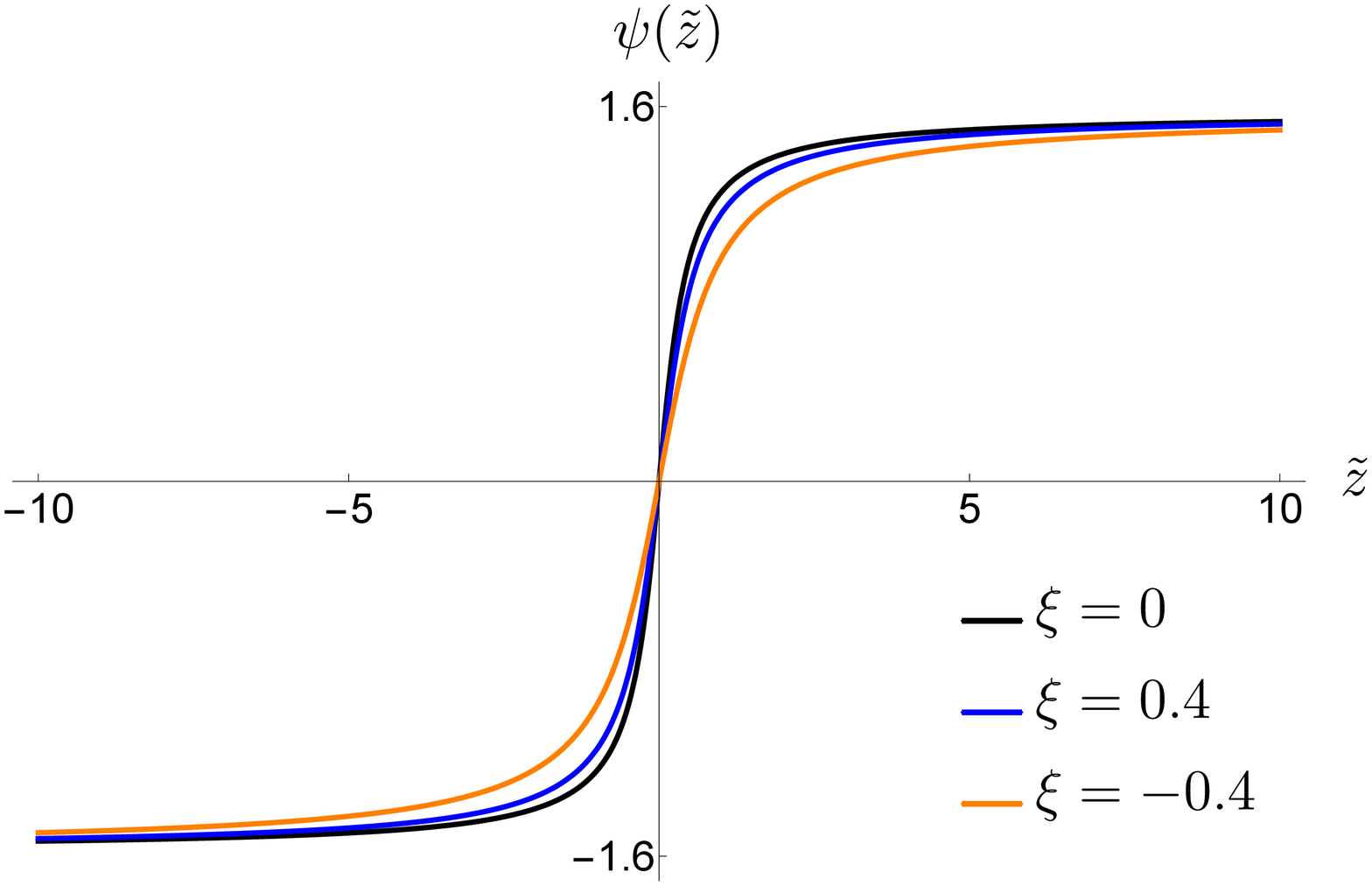}} \
\caption{(a) Intensity of the field $I ( \tilde{\rho}, \tilde{z} )$ (in units of $ I _{0} $) at its center (i.e. $\tilde{\rho} = 0$) as a function of the propagation distance $\tilde{z}$ for different values of $\xi$. (b) Rayleigh range $z _{R} (\xi )$ (in units of $z _{R} (0)$, the Rayleigh range in free space) as a function of the parameter $\xi$. (c) Gouy phase shift as a function of the propagation distance $\tilde{z}$ for different values of $\xi$.}
\end{figure}

Since the axial symmetry of the system still holds, even in the presence of a nontrivial background (which is along the radial direction), the propagated field will be axially symmetric too, which corresponds to $m=0$ in the Green's function (\ref{GreenFR}). The propagated field can be further evaluated by substituting the input field (\ref{hazgaussiano}) into Eq. (\ref{Prop}), with the Green's function given by Eq. (\ref{GreenFR}). Performing the changes of variables $\tilde{\rho} ^{\prime} = \rho ^{\prime} / \rho _{0}$ and $\tilde{k} = k \rho _{0}$, the field distribution at $z>0$ can be expressed in terms of the dimensionless coordinates $\tilde{\rho} = \rho  / \rho _{0} $ and $\tilde{z} = z / \rho _{0}$ as
\begin{align}
\phi _{\omega} ( \tilde{\rho}, \tilde{z} ) &= \frac{i \kappa}{1 - \xi} \, \sqrt{\frac{I _{0}}{8 \pi ( 1 - e ^{- 2 \kappa ^{2}} )}}  \int _{- \infty} ^{ \infty} \!\! d \tilde{k} \, e ^{i \tilde{k} \tilde{z} } \! \int _{0} ^{1} \tilde{\rho} ^{\prime} \left( \frac{ \tilde{\rho }}{ \tilde{\rho} ^{\prime} } \right) ^{\frac{\xi}{2(\xi - 1)}} J _{n _{0}} \left(  \sqrt{\frac{\Omega ^{2} - \tilde{k} ^{2}}{1 - \xi}} \, \tilde{\rho} _{<} \right)  H _{n _{0}} ^{(1)}  \left(  \sqrt{\frac{\Omega ^{2} - \tilde{k} ^{2}}{1 - \xi}} \, \tilde{\rho} _{>} \right) e^{- \tilde{\rho} ^{2} \kappa ^{2}}  d \tilde{\rho} ^{\prime} , \label{GaussianProp}
\end{align}
where $\tilde{\rho} _{<}$ ($\tilde{\rho} _{>}$) is the lesser (greater) between $\tilde{\rho}$ and $\tilde{\rho} ^{\prime}$, and $n _{0} = \frac{\xi}{2(1 - \xi )}$. Besides, we have defined $\Omega = \omega \rho _{0} / c$ and $\kappa = \rho _{0} / w _{0}$.The beam propagation, dictated by Eq. (\ref{GaussianProp}), is modified, relative to the standard scenario, by redefinitions on the frequency and wavenumber as well as by the presence of a nontrivial factor $(\tilde{\rho } /\tilde{\rho} ^{\prime}) ^{\frac{\xi}{2(\xi - 1)}}$. These integrals cannot be evaluated analytically in terms of known special functions, and hence we proceed to numerically evaluate it. Figure \ref{DensityG1} shows the field intensity profile $I ( \tilde{\rho}, \tilde{z} ) = \vert \phi _{\omega} ( \tilde{\rho}, \tilde{z} )  \vert ^{2} $ normalized in units of $I_{0}$ as a function of the dimensionless coordinates $\tilde{\rho}$ and $\tilde{z}$ for different values of the parameter $\xi$. In Fig. \ref{DensityG1ab} we show the difference between the intensities for the two particular values of the parameter $\xi=0$ and $\xi=0.4$, to be precise we plot $I( \tilde{\rho}, \tilde{z} )_{\xi = 0}-I( \tilde{\rho}, \tilde{z} )_{\xi = 0.4}$. At first sight, we observe that the Gaussian beam becomes more focused as the parameter $\xi$ is increased (as compared with the beam propagation in vacuum). To demonstrate this statement we require the following analysis regarding the evolution of the beam radius and beam waist as a function of $z$, moving away from a focal point. In Figs. \ref{dc1}-\ref{dc4} we present transverse cross-sections of the intensity profile as a function of the dimensionless propagation distance $\tilde{z}$ for different values of $\xi$, wherefrom we observe that the general Gaussian shape of the beam remains unaffected as it propagates. Also, we clearly see that the peak increases as the width becomes narrow for $\xi > 0$, as required by the total power conservation. This is the opposite behaviour to that in vacuum ($\xi = 0$), where the peak drops as the width broadens. This effect becomes more transparent in Fig. \ref{EjeRho0}, which displays the field intensity at its center (that is, at $\tilde{\rho} = 0$) as a function of the propagation distance $\tilde{z}$ for different values of $\xi$. Also, we can quantify the focalization of the beam by measuring the radius of the beam at a given propagation distance from the waist. To this end, we first recall that for a Gaussian beam propagating in free space, the Rayleigh range $z _{R} = \pi w _{0} ^{2} / \lambda$, with $\lambda$ the wavelength, gives the distance along the propagation direction of a beam from the waist to the place where the area of the cross section is doubled \cite{Svelto}. At a distance $z$ from the waist, the radius of the beam is $w (z) = w _{0} \sqrt{1+ (z / z _{R}) ^{2}}$. In the present case, we can estimate the Rayleigh range as a function of the parameter $\xi$ as follows. First, we evaluate the full width at half maximum (FWHM) as a function of $z$ (for different values of $\xi$) and then we use the relation for unapertured Gaussian beams
\begin{equation}
\frac{\textrm{FWHM}(z , \xi )}{\sqrt{2 \ln 2 }} = w _{0} \, \sqrt{1 + \left( \frac{z}{z _{R} (\xi ) } \right) ^{2}} \label{omegaZ}
\end{equation}
to fit our data and obtain $z _{R} (\xi )$, which we define as the Rayleigh range for the propagation of the Gaussian beam in this nontrivial background. Clearly, $z _{R} (0) = z _{R}$ is the Rayleigh range for propagation in free space. In Fig. \ref{ZRay1} we plot the normalized Rayleigh range $z _{R} (\xi ) / z _{R} (0)$ as a function of $\xi$. We observe that the Rayleigh range increases for $\xi > 0$, and decreases for $\xi < 0$. In other words, the beam becomes more focused for $\xi > 0$, while it rapidly drops out for $\xi < 0$. We emphasize that the above result is a rough estimation given that the Rayleigh range is defined by the relation in Eq. (\ref{omegaZ}), which corresponds to an unapertured Gaussian beam propagating in vacuum.

The Gouy phase shift is another of the important properties of Gaussian beams that deserves analysis due to its importance role in optics. The Gouy phase, given by $\psi (z) = \arctan (z / z _{R})$ in free space, is a phase advance gradually acquired by a Gaussian beam around the focal region during propagation \cite{Gouy, Simin}. It plays a role in the lateral trapping force at the focus of optical tweezers and leads to phase velocities that exceed the speed of light in vacuum \cite{Rohrbach}. The $\pi / 2$ Gouy phase shift has been directly observed in terahertz beams in a cylindrical focusing geometry \cite{McGowan}. In the presence of a nontrivial background, we expect a modification of the Gouy phase as a function of the parameter $\xi$. Having estimated the Rayleigh range $z _{R} (\xi )$ in this nontrivial background, we now can assess the Gouy phase by using the corresponding expression in free space $\psi (z)$ given above with the replacement $z _{R} \to z _{R} (\xi)$. Fig. \ref{Gouy123} shows the Gouy phase as a function of the propagation distance $\tilde{z}$ for different values of $\xi$. We observe that the parameter $\xi$ slows down the Gouy phase during propagation, i.e. it reaches the maximum of $\pi / 2$ over a longer distance as compared to that in free space.

\begin{figure}[H]
\centering
\subfloat[\label{f1}$z=0$]{%
\includegraphics[width=0.38\textwidth]{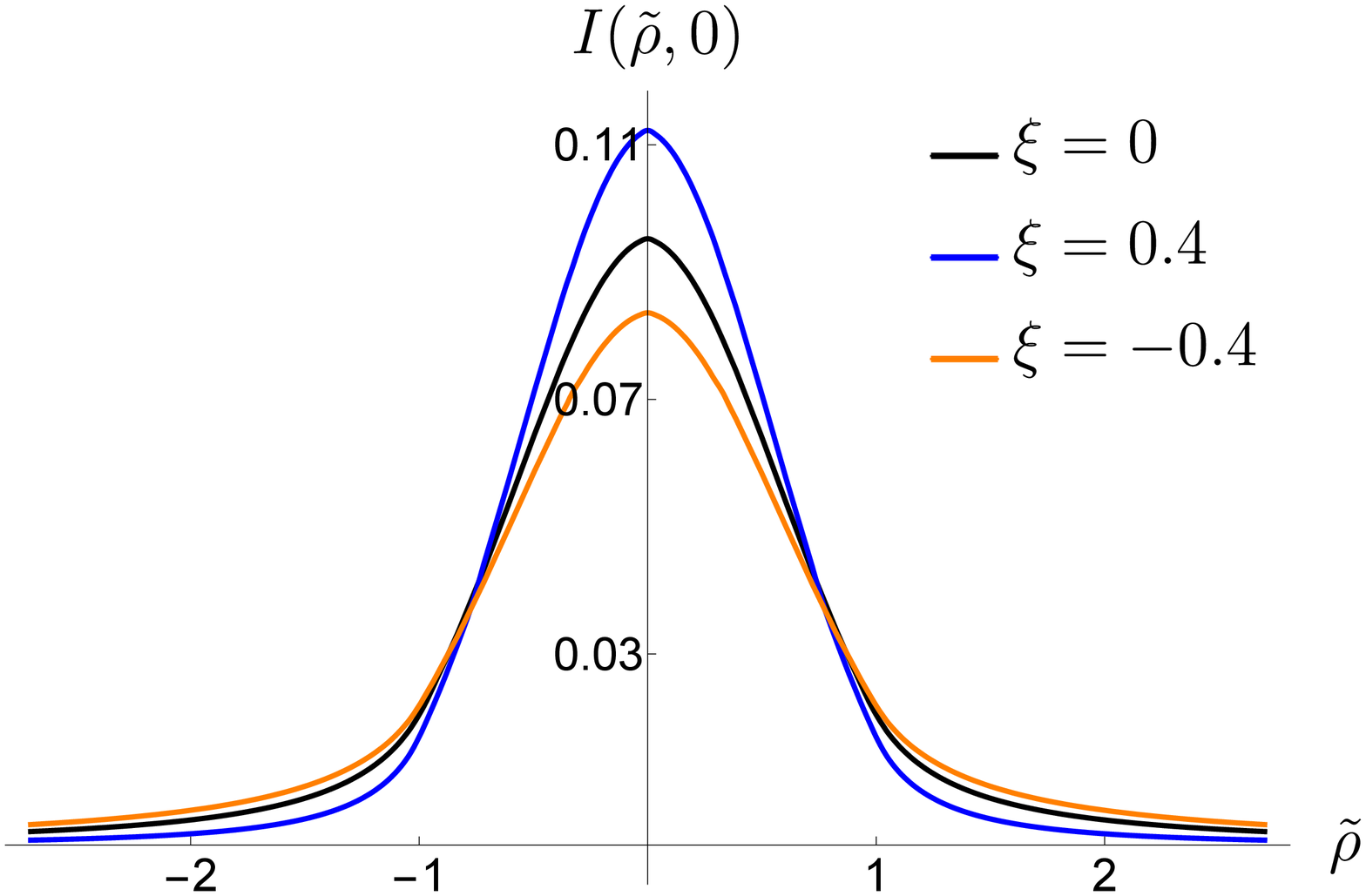}
}\hfill
\subfloat[\label{f2}$z=1.088$]{%
\includegraphics[width=0.38\textwidth]{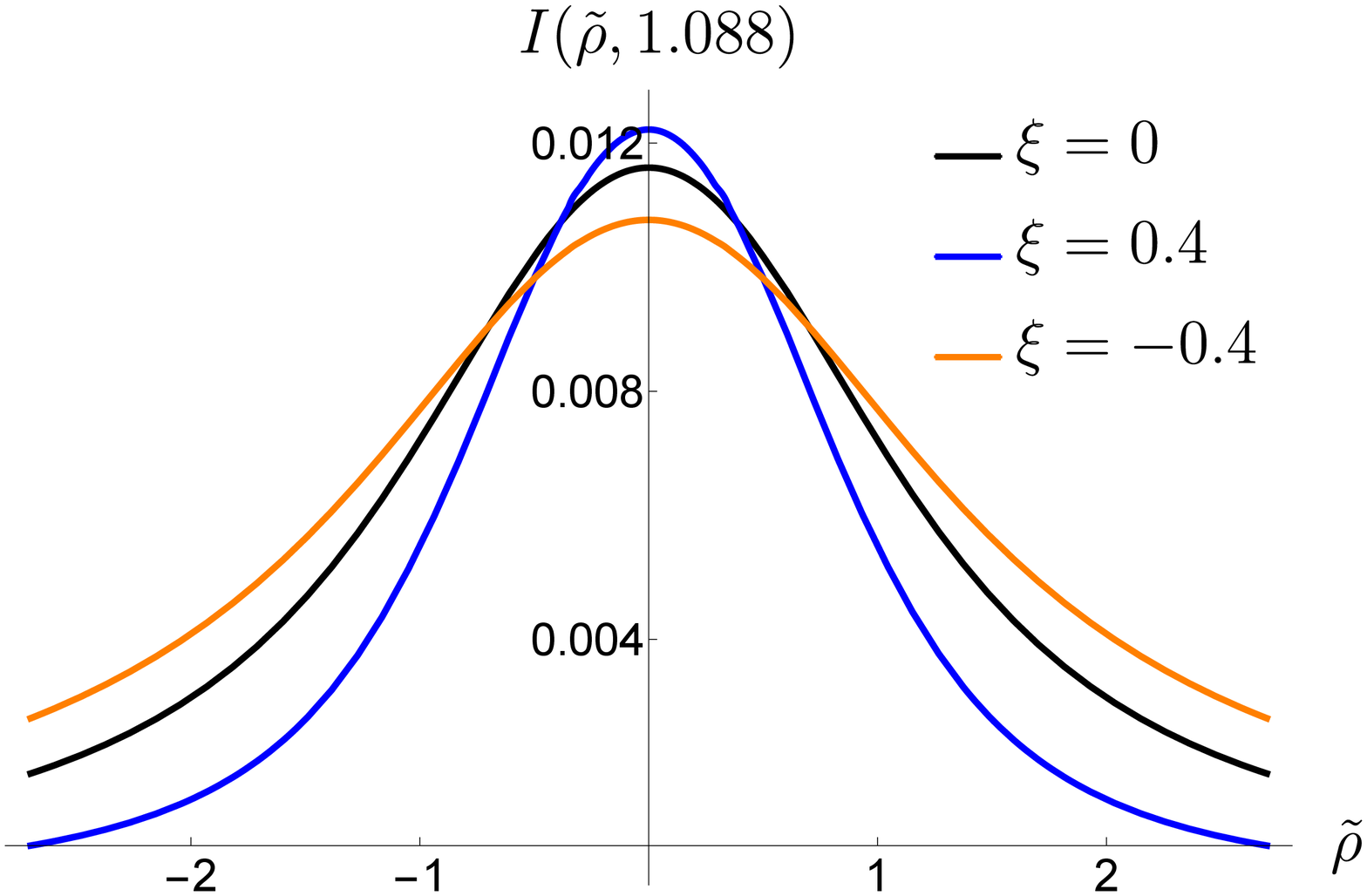}
}
\caption{Gauss profile (Intensity), considering $\xi$ negative}
\label{Fig5m}
\end{figure}

\subsection{Bessel beam}

Bessel beams represent a fundamental form of structured light \cite{mi11110997}. They are ideally propagation invariant and offer noteworthy features such as an exceptional depth-of-field and self-healing \cite{Durnin1987, Durnin1987.2}.  In complete analogy with ideal plane waves, an ideal Bessel beam has an infinite extent and energy, and thus only its finite aperture approximations are experimentally realizable \cite{LAPOINTE1992}. In practice, it has been used in optical trapping \cite{Skidanov2015, Khonina1999}, material processing \cite{Kuchmizhak2017, Khonina2016, Khonina2018}, optical coherence tomography \cite{FUJIMOTO2000}, superresolution \cite{QUABIS2000, Dorn2003, Sheppard2004}, sharp focusing \cite{Khonina2011, YouRong2013} and encryption in optical communications \cite{Chavez2018, KHONINA2019}. 

A Bessel beam is solution to the Helmholtz equation. The lowest mode of Bessel beams exhibits a zeroth-order Bessel function distribution in the perpendicular direction to the beam center line. In the present work we take the amplitude distribution of an apertured zeroth-order Bessel beam at $z=0$ the profile described by
\begin{equation}
\phi _{\omega} ( {\bf{r}} _{\perp} , 0 ) = \frac{1}{\rho _{0}} \sqrt{ \frac{I _{0}}{\pi [J _{0} ^{2} ( \alpha \rho _{0})  + J _{1} ^{2} ( \alpha \rho _{0} )]} }  J _{0} ( \alpha \rho) \, \Theta(\rho _{0} -  \rho  ) , \label{BesselInput}
\end{equation}
where $\alpha ^{-1}$ is an effective width, $\rho$ is the radial distance from the optical axis, and the Heaviside function expresses the finite aperture characteristic of the beam. The input (\ref{BesselInput}) is normalized to the power of the beam $I _{0}$, i.e. $ \int d ^{2} {\bf{r}} _{\perp} \, \vert \phi _{\omega} ( {\bf{r}} _{\perp} , 0 ) \vert ^{2} = I _{0}$.

\begin{figure}
\centering
\subfloat[\label{Ba1}$\xi=0$]{\includegraphics[width=0.42\textwidth]{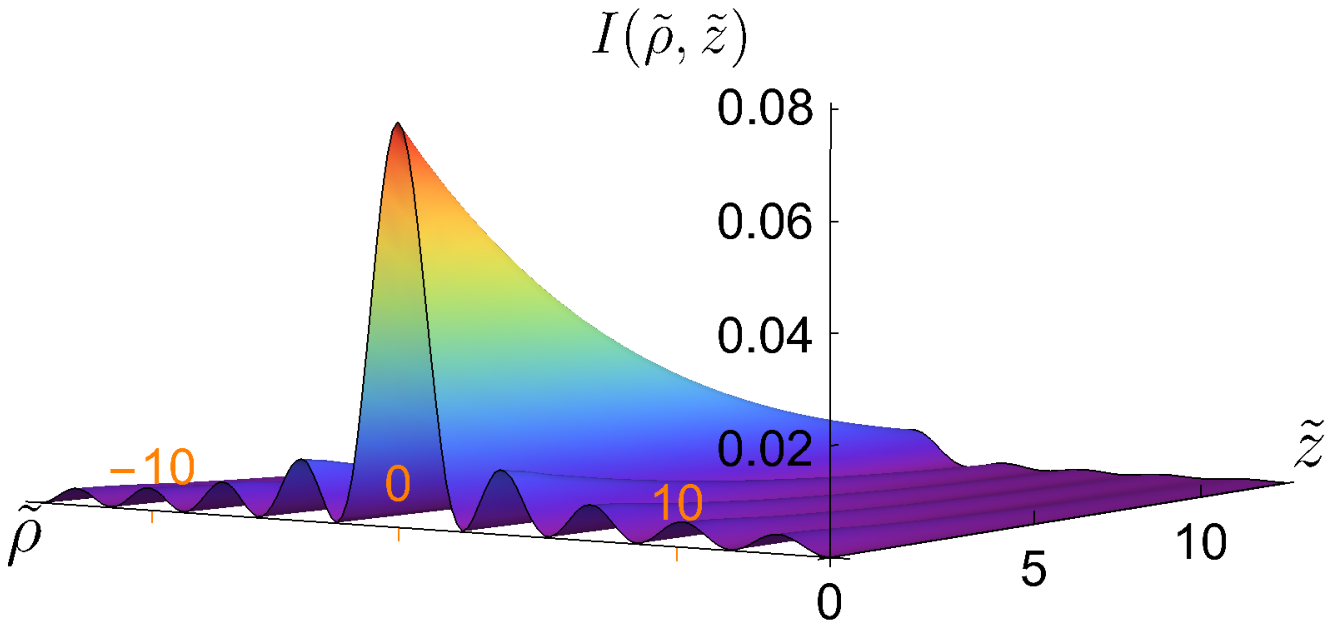}}  \hspace{1cm}
\subfloat[\label{Ba2}$\xi=0.04$]{\includegraphics[width=0.42\textwidth]{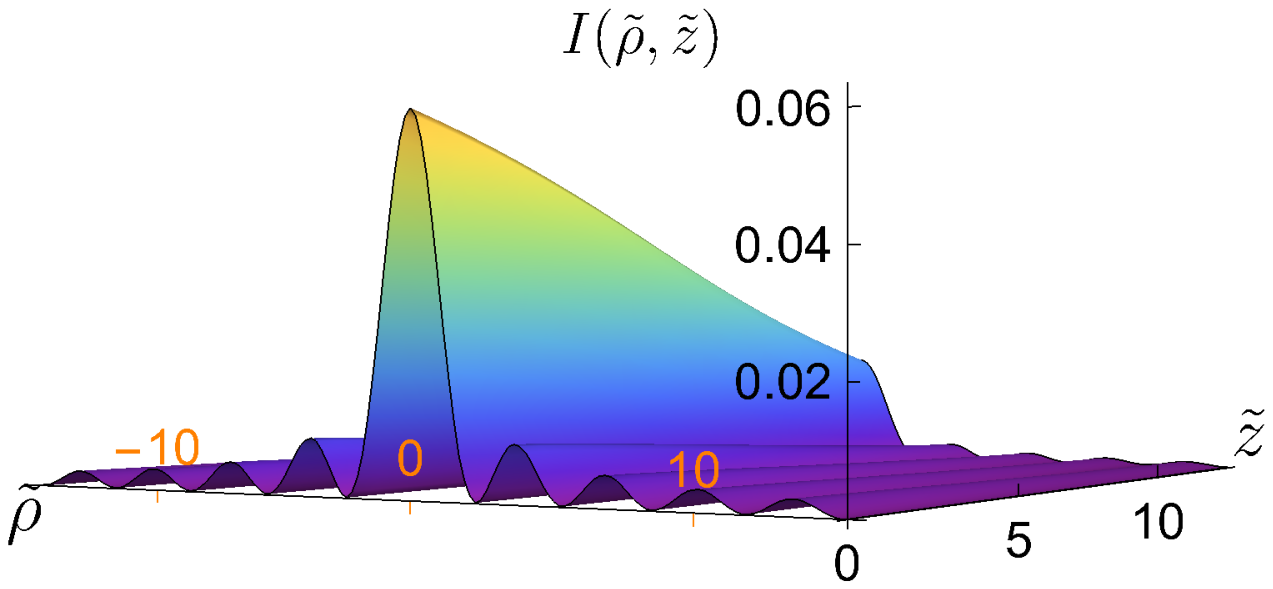}}
\caption{Apertured Bessel beam intensity $I ( \tilde{\rho}, \tilde{z} )$ (in units of $ I _{0} $) during propagation for (a) $\xi = 0$ and (b) $\xi = 0.04$.  }
\label{Bessel2a}
\end{figure}

\begin{figure}
\centering
\subfloat[\label{Bac1}$\xi=0$]{\includegraphics[width=0.36\textwidth]{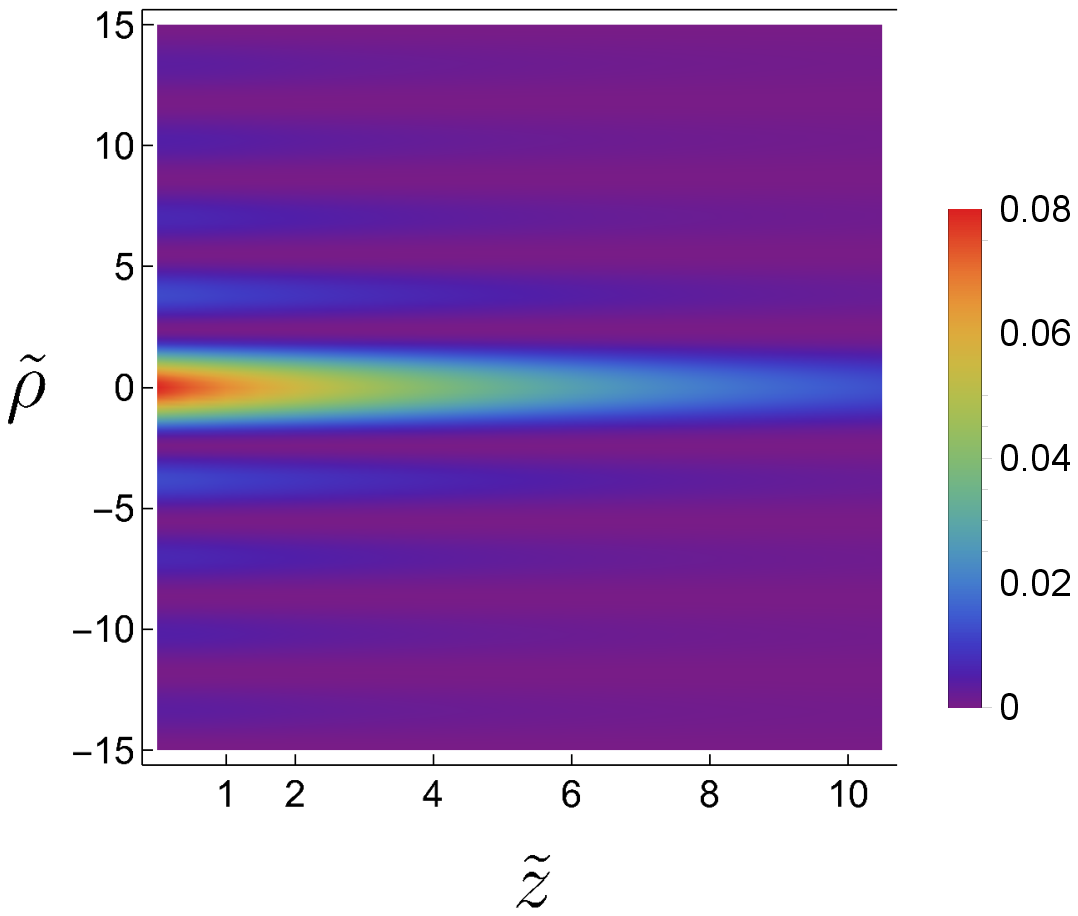}} \hspace{1cm}
\subfloat[\label{Bad1}$\xi=0.04$]{\includegraphics[width=0.36\textwidth]{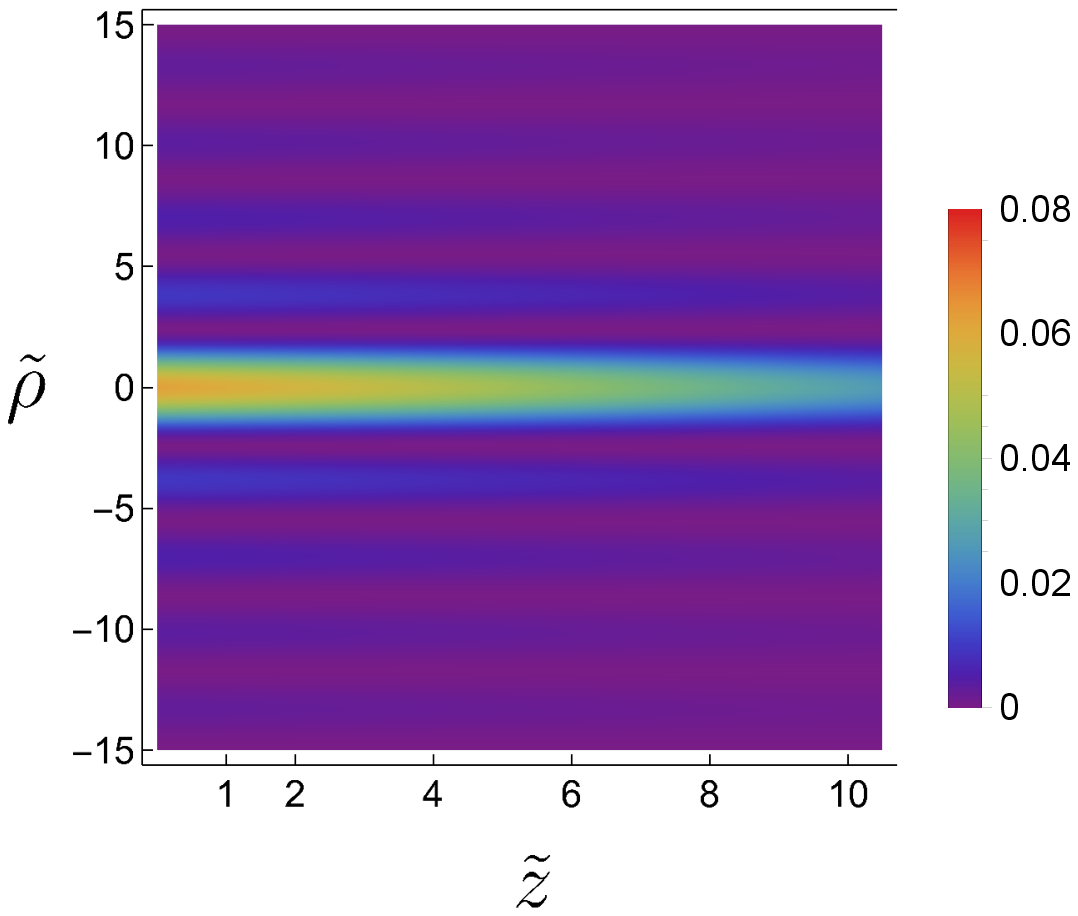}}
\caption{Density plot of the propagated apertured $J _{0}$ Bessel beam. }
\label{DensityPlotBessel}
\end{figure}

\begin{figure}
\centering
\subfloat[\label{Bd1}$z=0$]{\includegraphics[width=0.25\textwidth]{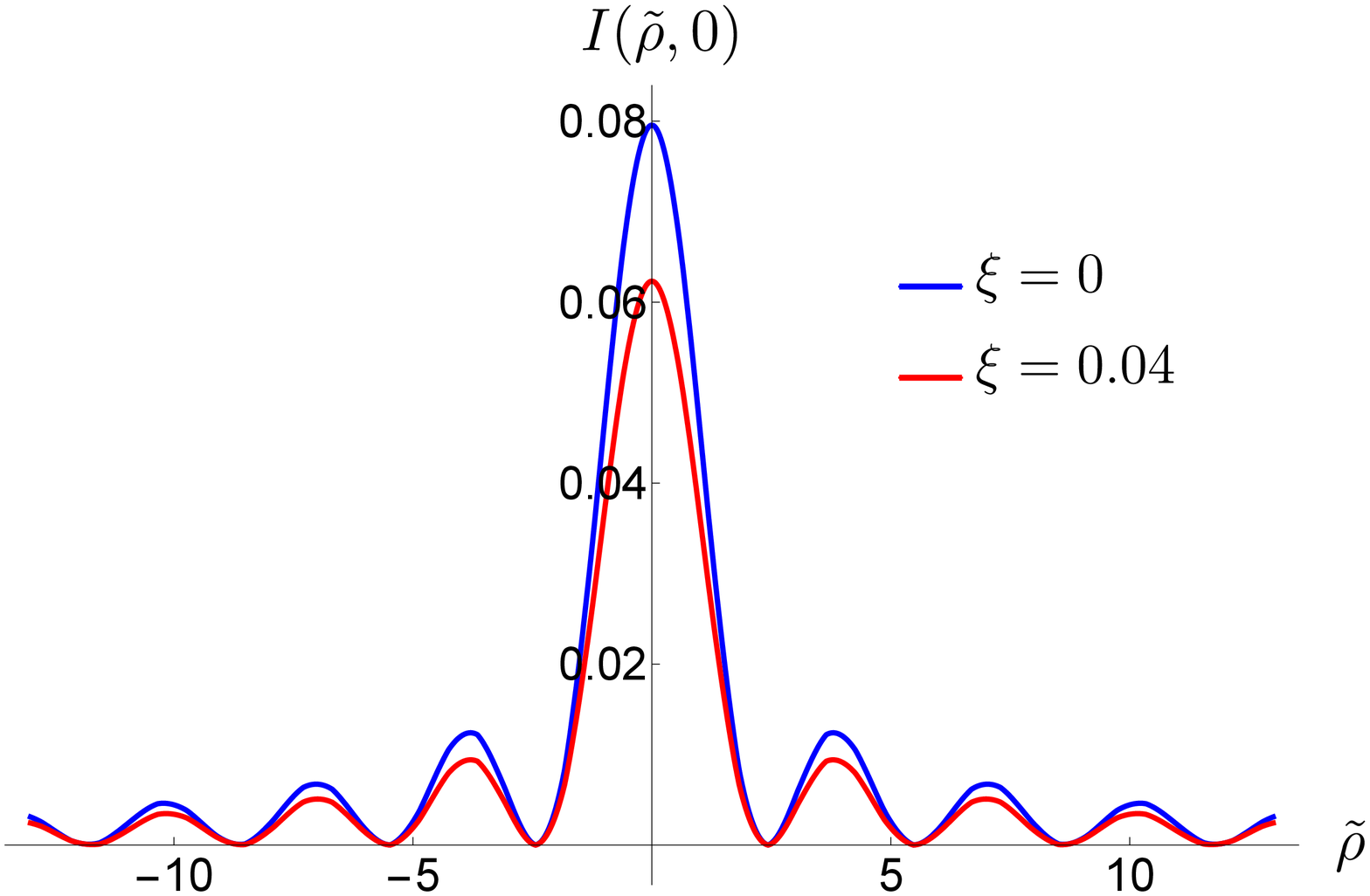}} \hfill
\subfloat[\label{Bd2}$z=3$]{\includegraphics[width=0.25\textwidth]{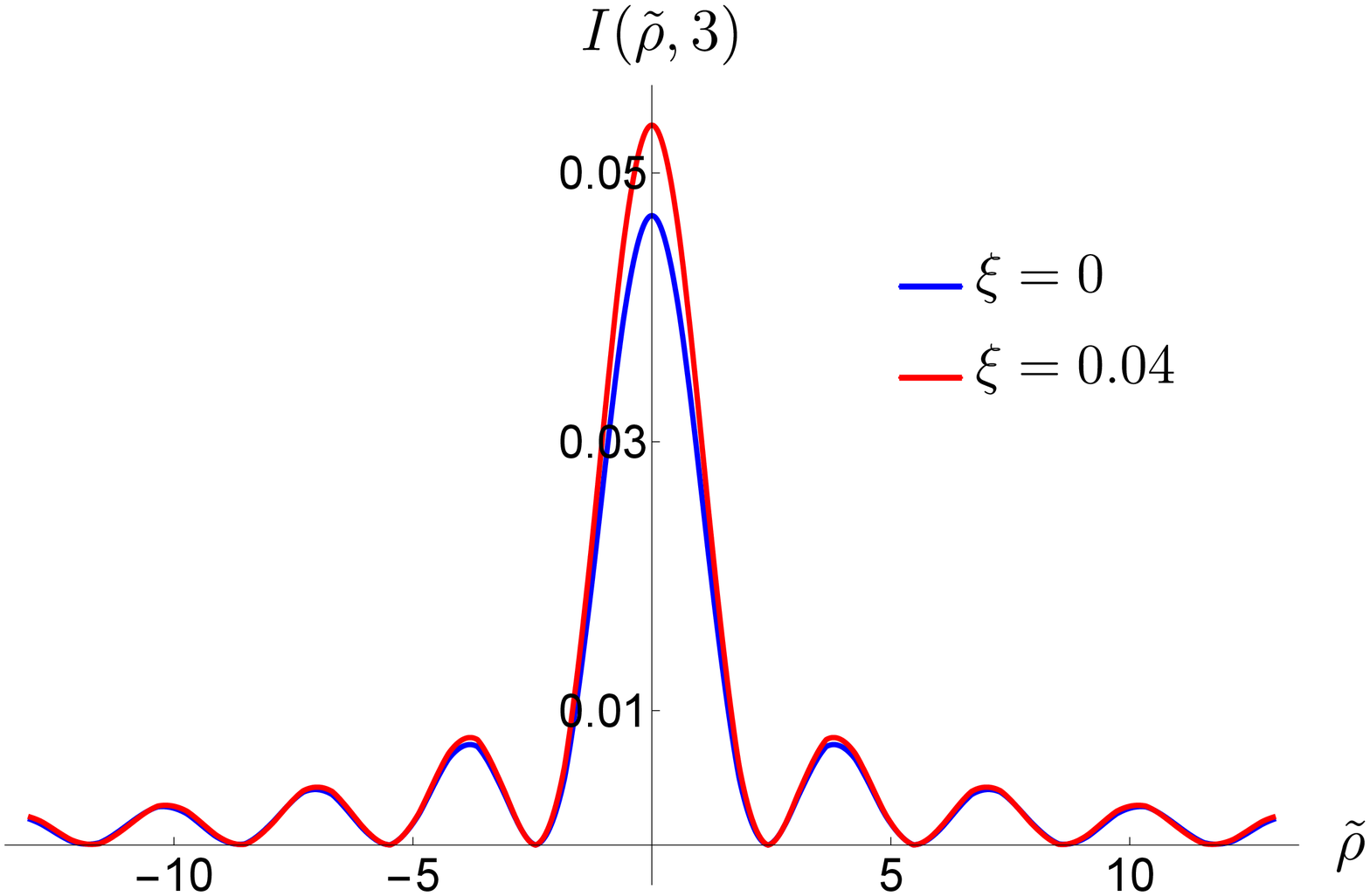}} \hfill
\subfloat[\label{Bd3}$z=8$]{\includegraphics[width=0.25\textwidth]{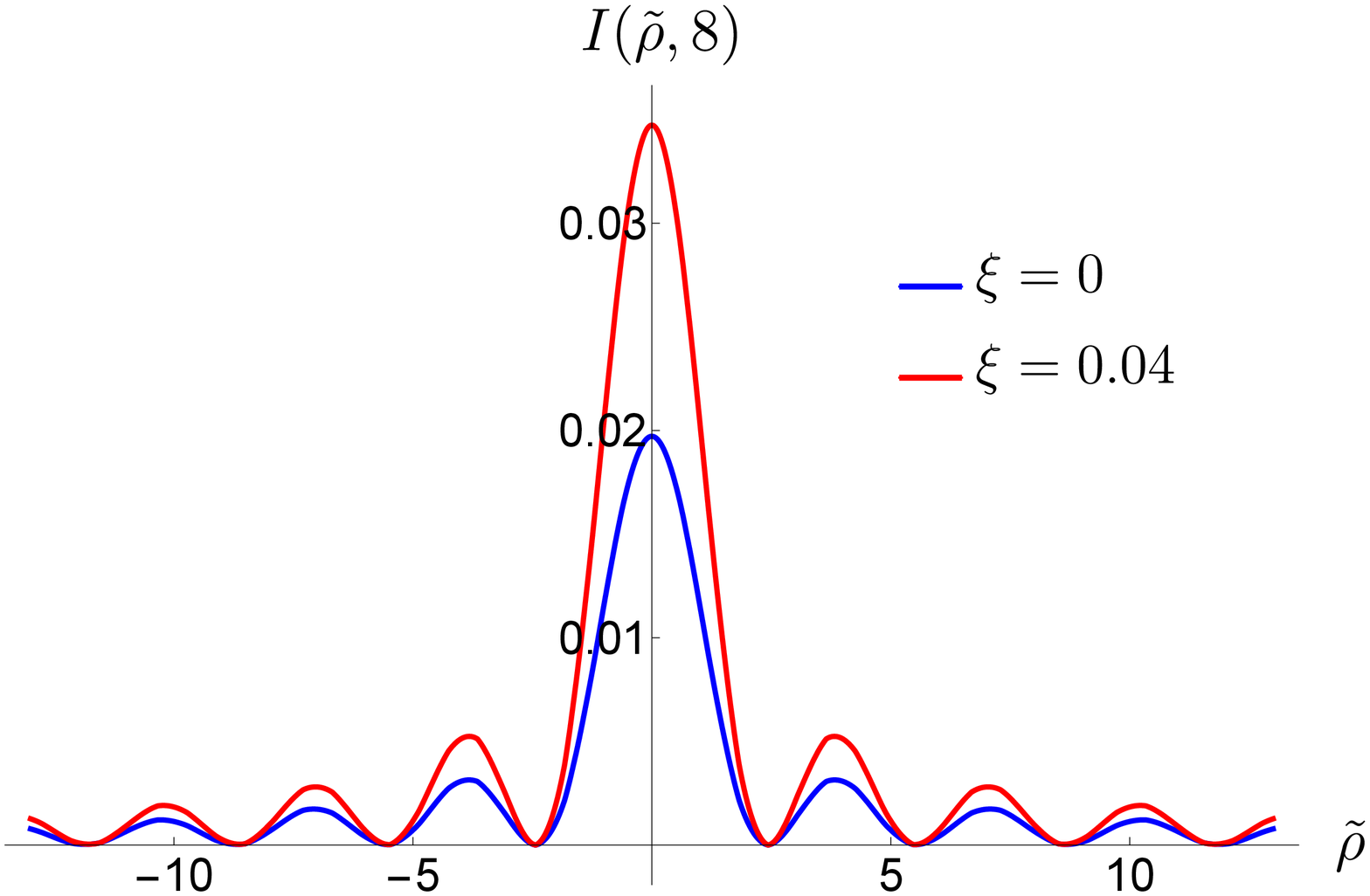}} \hfill \subfloat[\label{Bd4}$z=12$]{\includegraphics[width=0.25\textwidth]{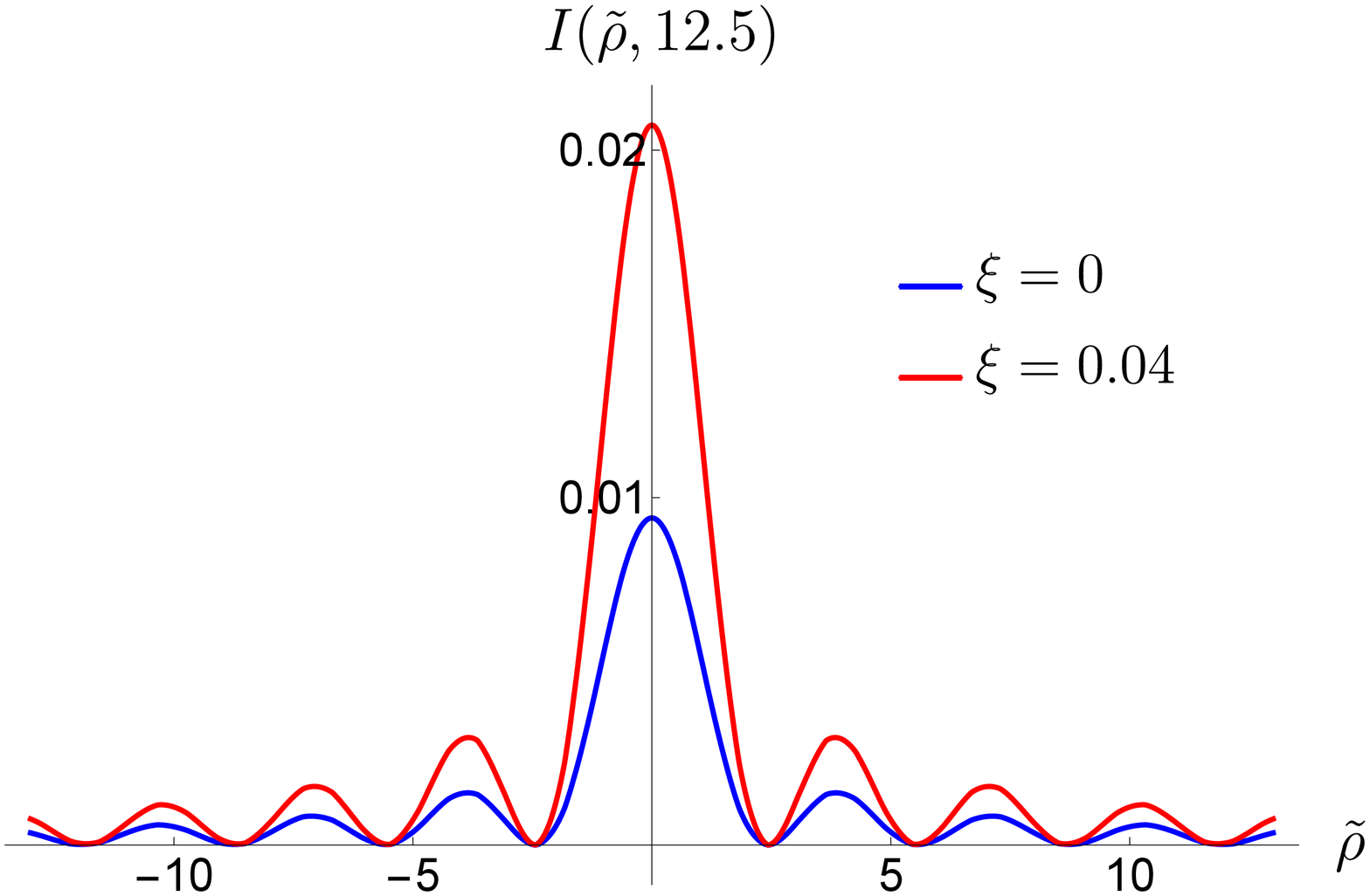}} \
\caption{Apertured Bessel beam intensity distribution $I ( \tilde{\rho}, \tilde{z} )$ (in units of $ I _{0} $) as a function of the dimensionless radial coordinate $\tilde{\rho}$, for $\xi = 0$ and $\xi = 0.04$, and different values of the dimensionless propagation distance $\tilde{z}$. }
\label{CrossSectionBessel}
\end{figure}

\begin{figure}
\includegraphics[width=0.3\textwidth]{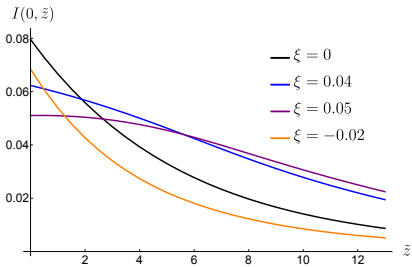}
\caption{ Intensity of the field $I ( \tilde{\rho}, \tilde{z} )$ (in units of $ I _{0} $) at its center (i.e. $\tilde{\rho} = 0$) as a function of the propagation distance $\tilde{z}$ for different values of $\xi$.  }
\label{BesselEjeZa}
\end{figure}

We now turn to propagate the input beam (\ref{BesselInput}). Due to the axial symmetry of the $J_{0}$ Bessel beam and the radially directed nontrivial background, we set $m = 0$ in Eq. (\ref{GreenFR}). Therefore, introducing the change of variables $\tilde{\rho} ^{\prime} = \rho ^{\prime} / \rho _{0}$ and $\tilde{k} = k \rho _{0}$, the propagated beam can be expressed in terms of the dimensionless variables  $\tilde{z} = z / \rho _{0}$ and $\tilde{\rho} = \rho / \rho _{0}$ as
\begin{align}
\phi _{\omega} ( \tilde{\rho}  , \tilde{z} ) &= \frac{i}{1 - \xi} \, \sqrt{\frac{I _{0}}{16 \pi [J _{0} ^{2} ( \tilde{\kappa} )  + J _{1} ^{2} ( \tilde{\kappa} )] }}  \int _{- \infty} ^{ \infty} d \tilde{k}  \, e ^{i \tilde{k}  \tilde{z}} \int _{0} ^{1} \,  \tilde{\rho} ^{\prime} \left( \frac{\tilde{\rho}}{\tilde{\rho} ^{\prime}  } \right) ^{\frac{\xi}{2(\xi - 1)}} J _{n _{0}} \left( \sqrt{\frac{ \tilde{\Omega} ^{2} - \tilde{\beta} ^{2}}{1 - \xi}} \tilde{\rho} _{<} \right) H _{n _{0}} ^{(1)} \left( \sqrt{\frac{ \tilde{\Omega} ^{2} - \tilde{\beta} ^{2}}{1 - \xi}} \tilde{\rho} _{>} \right)  J _{0} ( \tilde{\kappa} \tilde{\rho} ^{\prime}  ) d \tilde{\rho} ^{\prime}  , \label{BesselProp}
\end{align}
where $\tilde{\rho} _{<}$ ($\tilde{\rho} _{>}$) is the lesser (greater) between $ \tilde{\rho}$ and $ \tilde{\rho} ^{\prime} $. We have defined also $\tilde{\kappa} = \rho _{0} \alpha $ and  $\tilde{\Omega} = \omega \rho _{0} / c$. We now evaluate numerically the above integral. In Fig. \ref{Bessel2a} we present the field intensity profile $I (\tilde{\rho}, \tilde{z})= \vert \phi _{\omega} ( \tilde{\rho}  , \tilde{z} ) \vert ^{2}$ normalized in units of  $I _{0}$  as a function of the dimensionless coordinates $\tilde{\rho}$ and $\tilde{z}$ for different values of the parameter $\xi$. We observe that the Bessel beam propagates with preserved form, while also allowing for a concentrated beam profile. In contrast to the propagation of a $J _{0}$ Bessel beam in free space, in this case the field intensity becomes more focused as increasing the parameter $\xi$, which is similar to what occurs with the Gaussian beam (\ref{BesselProp}) in this nontrivial background. In the free-space case, the $J _{0}$ beam has a remarkably greater depth of field than the Gaussian, and it is due in large part to its energy distribution. Within the central maximum, the Bessel beam concentrates only 5\% of the total energy, which is sufficient to create a sharply defined central spot with an unchanging diameter over a large distance \cite{Durnin1987, Durnin1987.2}. Conversely, a Gaussian beam concentrates almost all its energy within the central spot, and this is why it spreads promptly. In the presence of a nontrivial background, as Fig. \ref{Bessel2a} suggests, the Bessel beam concentrates even more than in free space.  This indicates that the central spot of the beam concentrates most of the energy. To support this claim, in Fig. \ref{DensityPlotBessel} we present a density plot of the field intensity, wherefrom we observe that the beam focus is considerably enhanced by the parameter $\xi$. Also we can reach to this conclusion through the cross-sections of the density profile presented in Fig. \ref{CrossSectionBessel}. On the one hand, we observe the usual oscillatory behaviour with respect to the transverse coordinate $\tilde{\rho}$, together with the same pattern of the zeros of the intensity profile as that in free-space. In other words, the radial extension of the lobes is not affected by the nontrivial background, but the intensity does. Indeed, it is clear from Fig. \ref{CrossSectionBessel} that the energy of the central lobe (as well as the many side lobes), i.e. the area under the curve, for $\xi > 0$ is greater than the case of free-vacuum ($\xi = 0$), thus explaining such great depth of field. Finally, Fig. \ref{BesselEjeZa} displays the field intensity at its center (that is, at $\tilde{\rho}  = 0 $) as a function of the propagation distance $\tilde{\rho}$ for different values of $\xi$, wherefrom we observe again the enhanced focalization of the Bessel beam.

\section{Summary and discussion}
\label{Final1a}

In this paper we analysed the propagation of optical beams in a nontrivial spacetime background characterized by a constant four-vector $u ^{\mu}$, which can be interpreted whether as emerging from a weak gravitational field, from a nontrivial spacetime background arising from Lorentz symmetry breaking or from a general magnetoelectric and anistropic material media. To this end, we solved the modified wave equation by using Green's function techniques and applied the result to the propagation of finite apertured Gaussian and Bessel beams. Accordingly, we select the four-vector pointing along the radial direction (i.e. perpendicular to the beam propagation). We found that the nontrivial spacetime background affects the propagation of beams in different ways. The most salient is perhaps the enhancement of the Rayleigh range for beam propagation (as compared with the propagation in free space), i.e. they become more focused as the characteristic parameter $\xi$ (which controls the intensity of the background nontriviality) is increased. Conversely, the beams decay more rapidly as $\xi$ decreases.  

The field theory under consideration was originally conceived within the scalar sector of the Standard-Model Extension. However, since no deviations from Lorentz symmetry have been detected yet, the parameter $\xi $ is expected to be very small. Indeed, reference \cite{Kostelecky2011} provides a table of bounds on the Lorentz-violating coefficients in different sectors, including the photon sector which is the one relevant in this work. The stringentest bound reported for the CPT-even coefficients of the photon sector is with laser interferometry in gravitational wave detectors \cite{KOSTELECKY2016}, and is found to be of the order of $10 ^{-22}$, thus implying that the effects upon beam propagation predicted in this work are practically imperceptible. On the other hand, there are scenarios in which Lorentz symmetry is naturally broken, such as condensed matter systems. For example, the intrinsic anisotropy in a crystalline solid can play the role of the constant background, and this manifests in the optical and electronic properties of the material. In a similar vein, materials respond to strain, and it can be effectively described by fermions coupled to a weak gravitational field. In the case of moderate strain, the components of the strain tensor are approximately constants, and hence the macroscopic response is described by relativistic particles in a weak gravitational field (i.e. a constant background). In such cases, the relevant coefficients which quantify the symmetry-breaking are not necessarily small. In these scenarios, our results acquire relevance, since the modified wave equation could have different origins from a condensed matter perspective. 

The applications of optical beams, in particular Bessel beams, in nonlinear optics, has gained great interest in the last years because they can produce strongly peaked pump-intensity distribution over relatively long spans in nonlinear media. In many applications, such as laser materials processing or surgery, it is highly important to focus a laser beam down to the smallest spot possible to maximize intensity and minimize the heated area. In this regard various mechanisms of focusing medium nonlinearity and the longitudinal self-modulation of propagating beams have been investigated \cite{Porras2004, Jukna2014, Polesana2005, GADONAS2001}. For example, the effect of a Kerr nonlinearity (modelled by the nonlinear Schr\"{o}dinger equation) upon the propagation of non-diffractive Bessel beams  has been studied in Ref. \cite{JOHANNISSON2003107}, and the main finding is that the nonlinearity is focusing or defocusing depending on the sign of the nonlinear index coefficient. In this work we showed that a similar effect appears when optical beams propagate through nontrivial spacetime backgrounds, in which case the sign of the parameter $\xi$ determines the focusing or defocusing of the beam.

We close by commenting on possible realizations of the predicted effects with material media. As discussed in Section \ref{Model1}, the electrodynamics model behind the scalar field theory considered in this work arises from the photon sector of the Standard-Model Extension, and may be understood as an anisotropic and magnetoelectric material media, for example, topological phases and multiferroics. The difference between them is the origin of the magnetoelectricity. While the magnetoelectric coupling of the former lies in the nontrivial topology of the underlying band structure, in the latter it arises from the coupling between spins and electric dipoles in the material. The most promising materials are perhaps multifferroics, since there is an essential ingredient required to observe optical beam propagation in the media: optical transparency. To the best of our knowledge, the optical transparency of topological materials has not been investigated yet. However, topological insulators can be understood as a bulk dielectric with a surface Hall effect, and therefore the optical transparent region may corresponds to the trivial regions of the bulk dielectric.  Of course, this requires a formal study. On the other hand, the optical transparency of light in multiferroic materials, such as Ca$_2$CoSi$_2$O$_7$, Sr$_2$CoSi$_2$O$_7$ and Ba$_2$CoGe$_2$O$_7$, has been extensively studied both from the theoretical and experimental sides \cite{PhysRevLett.115.267207, Kezsmarki2014}. Magnetoelectric multifferoics are simultaneously both ferromagnetic and ferroelectric, and the dynamical or optical magnetoelectricity appears due to the coupling of the coexisting order parameters. When the optical magnetoelectric effect is strong enough, the aforementioned materials become fully transparent for a given propagation direction, while they absorb light travelling in the opposite direction. Therefore, they represent a good test bed for the predictions presented in this paper. The next obstacle is related to the geometry of the material. We must guarantee that the material exhibits an axially symmetric magnetoelectric tensor. This is of course a great experimental challenge. Along this line, recent studies report cylindrically symmetric optical transparent magnetoelectric materials composed by magnetostrictive Fe$_{72.5}$Si$_{12.5}$B$_{15}$ microwires and piezoelectric poly(vinylidene fluoride-trifluoroethylene) \cite{aelm.201900280}.  All in all,  our predictions are theoretical for the time being,  but we expect that with the recent advances in optical transparent magnetoelectric materials,  a properly engineered material (on demand) may simulate the properties of the cylindrically-symmetric nontrivial spacetime background considered in this work,  and then observe the reported focalization properties.

\acknowledgements{C.A.E. and M.M. acknowledge support from PAPIIT UNAM project No. IN109321. A.M.-R. has been partially supported by DGAPA-UNAM Project No. IA102722 and by Project CONACyT (M\'{e}xico) No. 428214. }

\bibliography{references.bib}
\end{document}